\documentclass[preprint]{aastex} 

\newcommand{\be}{\begin{equation}}
\newcommand{\ee}{\end{equation}}
\newcommand{\np}{ {\cal N} }

\newcommand{\mbar}{ {\langle m \rangle} } 
\newcommand{\taudamp}{ { \tau_{\rm damp}}} 
\newcommand{\taudisk}{ { \tau_{\rm disk}}} 

\begin{document}

\title{MIGRATION AND DYNAMICAL RELAXATION \\
IN CROWDED SYSTEMS OF GIANT PLANETS} 

\medskip
\author{Fred C. Adams$^{1,2}$ and Gregory Laughlin$^3$}
\bigskip 

\affil{$^1$Michigan Center for Theoretical Physics \\ 
Physics Department, University of Michigan, Ann Arbor, MI 48109}

\affil{$^2$Astronomy Department, University of Michigan, Ann Arbor, MI 48109}

\affil{$^3$Lick Observatory, University of California, Santa Cruz, CA 95064} 

\begin{abstract} 

This paper explores the intermediate-time dynamics of newly formed
solar systems with a focus on possible mechanisms for planetary
migration. We consider two limiting corners of the available parameter
space -- crowded systems containing $\np=10$ giant planets in the
outer solar system, and solar systems with $\np=2$ planets that are
tidally interacting with a circumstellar disk.  Crowded planetary
systems can be formed in accumulation scenarios -- if the disk is
metal rich and has large mass -- and through gravitational
instabilities. The planetary system adjusts itself toward stability by
spreading out, ejecting planets, and sending bodies into the central
star. For a given set of initial conditions, dynamical relaxation
leads to a well-defined distribution of possible solar systems. For
each class of initial conditions, we perform large numbers (hundreds
to thousands) of N-body simulations to obtain a statistical
description of the possible outcomes.  For $\np=10$ planet systems, we
consider several different planetary mass distributions; we also
perform secondary sets of simulations to explore chaotic behavior and
longer term dynamical evolution. For systems with 10 planets initially
populating the radial range 5 AU $\le a \le$ 30 AU, these scattering
processes naturally produce planetary orbits with $a\sim1$ AU and the
full range of possible eccentricity ($0 \le \epsilon \le 1$). Shorter
period orbits (smaller $a$) are difficult to achieve. To account for
the observed eccentric giant planets, we also explore a mechanism that
combines dynamical scattering and tidal interactions with a
circumstellar disk. This combined model naturally produces the
observed range of semi-major axis $a$ and eccentricity $\epsilon$. We
discuss the relative merits of the different migration mechanisms for
producing the observed eccentric giant planets.

\end{abstract} 

\medskip 
$\,$  


\newpage 

\section{INTRODUCTION}           \label{sec:intro} 

The past decade has witnessed a revolution in the study of planetary
systems, with over one hundred extrasolar planets discovered thus
far. The initial discoveries (Mayor and Queloz 1995; Marcy and Butler
1996) showed an unexpected feature -- namely that the orbital
parameters of the newly discovered planets were markedly different
from those of the planets in our solar system. Many of the giant
planets are found in short period orbits ($P_{\rm orb} \approx 4$
days) while others display longer orbits of high eccentricity (0 $\le
\epsilon \le$ 0.92). Subsequent discoveries (e.g., Marcy and Butler
1998, 2000; Hatzes et al. 2000; Perryman 2000) have shown that such
planetary systems are relatively common.  Approximately 8 percent of
the stars in the solar neighborhood have giant planets in tight orbits
with $a < 3$ AU. Hot Jupiters -- giant planets in $\sim4$ day orbits
-- account for one eighth of the observed sample of planets (1 percent
of the stars). The remaining seven eighths of the observational sample
(7 percent of the stars) have eccentric giant planets. This observed
population of planets shows an apparent deficit of orbits with periods
$P$ in the range 10 $< P <$ 100 days. If this observational trend
holds up, it may suggest that different migration (or formation)
mechanisms are at work for the hot Jupiters ($P \sim 4$ days) and the
eccentric giants ($P \ge 100$ days). In any case, an explanation for
the origin of these orbits poses an important astronomical problem.

Theories of planet formation come in two main varieties. The leading
theory, at least for our own solar system, holds that planets form
``from the bottom up'' through the gradual accumulation of
planetessimals (see the review of Lissauer 1993). These rocky building
blocks collect into larger entities until their gravitational
influence is strong enough to accrete gas from the surrounding
nebula. In circumstellar disks with sufficiently high mass, the ascent
from planetessimals to giant planets can occur rapidly and can lead to
crowded systems (Lissauer 1987; Lin and Ida 1997; see also Levison,
Lissauer, and Duncan 1998). The majority of the newly discovered
planetary systems orbit stars with high metallicity (e.g., Gonzalez
1997, Laughlin 2000), which supports the production of more rocky
material and enhances the rapid assembly of giant planets.  The
alternate theory holds that planets form ``from the top down'' through
the action of gravitational instabilities in the circumstellar disk
(e.g., Cameron 1978; Boss 2000). Under ideal conditions, gravitational
instabilities can grow on a time scale comparable to the orbital
period of the outer disk edge (Adams, Ruden, and Shu 1989). For disks
with radius $r_{\rm disk} \approx$ 30 AU, for example, the instability
time scale is thousands of years, much shorter than the relaxation
time of the system (see below).  As a result, giant planet formation
can occur even more rapidly through this channel.

In both scenarios outlined above, planet formation can proceed -- at
least in principle -- faster than dynamical relaxation of the newly
formed system.  The initial states for the planetary systems are not,
in general, dynamically stable over much longer time intervals. The
astronomical motivation for this present study is to explore the
dynamics of these crowded planetary systems in greater detail. In
particular, we can find the odds of obtaining high eccentricity
planets with $a \sim 1$ AU (like many of those observed).

This study has another objective. Solar system formation is likely to
be chaotic -- in the technical sense. Due to sensitive dependence on
the starting conditions, the result of any particular realization of
solar system formation cannot be described in terms of a single
outcome. Instead, a set of physically equivalent starting conditions
will generally display a full distribution of outcomes. In this study,
we illustrate this behavior explicitly by calculating the distribution
of outcomes for each chosen set of initial conditions.

In addition to the astronomical applications, this study provides an
interesting problem in dynamics. These planetary systems have their
gravitational potential dominated by the central star and they begin
with their primary motions as orbits about the central mass. As a
result, the relaxation of a planetary system will be somewhat
different from that of a stellar system (such as a globular cluster --
see Binney and Tremaine 1987).

This research builds on previous work. Dynamical instabilities
involving two giant planets have been explored extensively through
numerical simulations (Ford, Havlickova, and Rasio 2001). This work
showed that interactions between only two planets (with equal mass)
cannot reproduce the orbital characteristics of the observed
extrasolar planets (see also Rasio and Ford 1996; Weidenschilling and
Marzari 1996); similarly, a single close encounter between two planets
cannot explain the observed orbits (Katz 1997).  A more recent study
considers dynamical instabilities involving three Jupiter-mass planets
(Marzari and Weidenschilling 2002) and compares favorably with the
observations (see also Ford, Radio, and Yu 2002 for the case of
unequal mass planets).  Dynamical relaxation in larger systems of
extrasolar planets has been considered by Papaloizou and Terquem
(2001; see also Terquem and Papaloizou 2002; Lin and Ida 1997;
Chambers, Wetherill, and Boss 1996). As noted by many previous
authors, the parameter space available to multiple planet solar
systems is enormous. The dynamical relaxation portion of this study
extends previous work by providing a systematic exploration of one
region of parameter space -- that containing $\np$ = 10 planets within
30 AU of the central star. In order to obtain statistically meaningful
results, this work employs many realizations (typically, $N$ = 100) of
any given starting condition and determines the distribution of
possible outcomes.

Many previous authors have also considered tidal torques exerted on
planets by circumstellar disks. Migration was anticipated long before
extrasolar planets were detected (e.g., Goldreich and Tremaine 1980;
Lin and Papaloizou 1993). With the discovery of extrasolar planets in
short period orbits, many studies of migration have been carried out
(e.g., Lin, Bodenheimer, and Richardson 1996; Trilling et al. 1998;
Nelson et al. 2000). These studies generally consider only one planet
at a time; however, two planet systems have recently been considered
(Lee and Peale 2002; Murray, Paskowitz, and Holman 2002).  In this
work, we extend these calculations to include instabilities between
multiple planets in the presence of tidal torques from a surrounding
disk. The simultaneous action of both dynamical scattering and viscous
torquing allows for a wider variety of behavior and final system
properties. The tidal torques are efficient at moving planets inward,
while the scattering interactions are effective at increasing orbital
eccentricity. The combination naturally produces planetary orbits with
small semi-major axis $a$ and large eccentricity $\epsilon$, much like
some of the observed eccentric giant planets.

This paper is organized as follows. In \S 2, we describe the numerical
treatment and the collections of initial conditions used for the both
the dynamical relaxation experiments and the disk-torquing simulations.
The basic dynamical results are presented in \S 3, including a
specification of the distributions of final system properties. We also
demonstrate the chaotic nature of the dynamics and find the dependence
of the time scales on the planetary masses and other parameters of the
problem.  In \S 4, we present a complementary set of simulations that
include scattering of only two planets, but include the tidal
interaction of the outer planet with a circumstellar disk. We conclude
with a summary and discussion of our results in \S 5.

\section{METHODS AND INITIAL CONDITIONS}    \label{sec:init} 

The focus of this investigation is to perform numerical simulations of
nascent planetary systems. The numerical experiments were carried out
by using the {\sl Mercury 6} integration package (Chambers 1999),
which provides general-purpose software for N-body integrations. Since
the code was designed to calculate the orbital evolution of objects
moving in the gravitational field of a central star, it is well-suited
for this purpose. In order to maintain sufficient accuracy, we use the
Bulirsch-Stoer (BS) integration scheme for all of the simulations
presented herein. Although somewhat slower than other computational
options, the BS scheme maintains greater accuracy during close
encounters, which drive the evolution of these planetary systems.
Specifically, we use an intergration accuracy parameter of $10^{-11}$
which sets the maximum fractional error per time step. For the course
of one million years integrations (see below), the accumulated
fractional error in the energy $\Delta E/E$ is typically a few parts
per million. Angular momentum is conserved to greater accuracy, with a
typical fractional error in $\Delta J/J$ of only a few parts per
billion.

In the first phase of this investigation, we set up a series of
simulations with the following properties. Each system begins with
$\np$ = 10 planets orbiting a star with mass $M_\ast$ = 1.0 $M_\odot$.
We perform four ensembles of $N$ = 100 simulations, where each set
uses a particular distribution to specify the planetary masses (see
below). At the start of each simulation, the planets are placed on
circular orbits, with the logarithm of the orbital radius chosen
randomly over the range corresponding to 5 AU $\le r \le$ 30 AU (the
range of semi-major axes sampled by giant planets in our solar
system).  The initial velocities are chosen to be those appropriate
for circular orbits at the given radius. The angular location of the
planets is random. Furthermore, the planets are displaced above or
below the orbital plane by a small amount, a randomly chosen distance
between zero and a ``scale height'' defined to be $H = 0.05 r$. With
this set of initial conditions, the planets are then integrated for a
time interval of one million years.

For our four initial ensembles of simulations, we adopt the following
mass distributions: [A] All 10 planets have equal mass, where $m_P$ =
$m_J$ = 0.001 $M_\odot$.  [B] All 10 planets have equal mass, where
$m_P$ = $2 m_J$ = 0.002 $M_\odot$.  [C] The 10 planets have masses
drawn from a random (uniform) distribution over the mass range $0 \le
m_P \le 4 m_J$ = 0.004 $M_\odot$.  [D] The 10 planets have masses that
are uniformly distributed in $\log m$ and are chosen from the range
$-1 \le \log_{10} [m_P/m_J] \le 1$. In all cases, the planetary masses
are drawn independently. For each mass distribution, we perform
$N=100$ separate numerical integrations ($N$=100 different
realizations of the same class of initial conditions). The resulting
ensembles of solar systems (100 for each mass distribution) can be
characterized in a statistically significant manner. For the two of
the mass distributions (the uniform random and logarithmic random
cases), we follow up these initial numerical experiments with a
smaller number ($N$ = 25) of longer term integrations. Finally, we
consider another mass distribution that contains one high mass planet
($m_P$ = $m_J$) and 19 smaller planets with mass in the range 
0 $\le m_P \le$ 0.5 $m_J$ (see \S 3.5). 

In this initial set of simulations, we ignore the possibility of
planets merging. For two planet systems, the effects of merging have
been clearly delineated (Ford et al. 2001). When the parameter $b_M
\equiv (r_P/r_J) (a/5 {\rm AU})^{-1}$ is less then unity, collisions
between planets are relatively unimportant (see Fig. 7 of Ford et
al. 2001).  As a result, if the planets have merging cross sections
that are comparable to the physical size of present-day giant planets,
merging can be safely neglected for the regime of parameter space
sampled by our simulations. As a benchmark, in \S 4 we consider tidal
torques acting on pairs of planets, which migrate inwards to small
semi-major axes; collisions are included in these simulations, where
the effective planetary radius $r_P \approx 2 r_J$. The results of 
\S 4 show that collisions take place in less than 2 percent of the
systems and that the merging planets typically have radial locations
$r \sim 1 - 3$ AU.  Furthermore, the results of \S 3 show that when
planets are scattered inward to $a\sim1$ AU, they are typically well
isolated, with the semi-major axis of the next closest planet a factor
of 15 -- 20 farther larger. If the planets form through gravitational
instability, however, the cross sections can be much larger and
merging can become significant (see, e.g., Lin and Ida 1997). Notice
also that in the absence of collisions, we could rescale our
simulations to study starting conditions inside our chosen boundary at
5 AU. With $r_P \sim r_J$, we can only rescale the inner boundary to
about 1 AU before collisions start to become important.

In \S 4, we consider smaller systems with only two planets, but allow
the outer planet to be acted upon by an external torque from a
surrounding circumstellar disk. These integrations are carried out
using a BS scheme adopted from codes developed previously to study
solar system scattering cross sections (Laughlin and Adams 2000; Adams
and Laughlin 2001). This code is supplemented with subroutines form
the Mercury 6 integration package (Chambers 1999). By using a code
that is explicitly optimized for the few-body problem in the main
integration, in conjunction with separately optimized subroutines for
evaluating orbital elements, the code maintains both high accuracy and
high speed. As before, we perform a large number (hundreds) of
realizations of the problem to obtain a good determination of the
distributions of possible outcomes.

\section{RESULTS}            \label{sec:results} 

For each of the four distributions outlined above, we perform $N=100$
simulations and use the results to define the characteristic time
scales (\S 3.1) and build up a statistical description of the outcomes
(\S 3.2). We are particularly interested in the properties of the
innermost surviving planets in these systems (\S 3.3), as these
survivers may explain the observed eccentric giant planets in the
observational sample.  With these results in hand, we also perform
secondary sets of simulations to explore particular dynamical issues,
including the chaotic nature of the dynamics (\S 3.4), scattering into
resonant configurations (\S 3.5), the longer term dynamical evolution
of the systems (\S 3.6), and more extreme planetary mass distributions
-- the action of many small planets on one large planet (also \S 3.6).

\subsection{Time Scales} 

Crowded planetary systems, and the numerical simulations presented
herein, display two important time scales.  The first is the decay
time, which is the time required for a solar system to decay by either
ejecting a planet or sending a planet into the central star. The
second time scale is the evolution time, i.e., the time over which the
solar system adjusts itself to stability.

The statistics of solar system decay can be considered analogous to
that used to describe the decay of atomic nuclei. The systems start
with 10 planets and then decay into daughter systems with 9
planets. In this context, the decay has two channels, either outright
ejection or accretion of the planet by the central star. In this
context, we consider both channels to be different modes of the same
effect. In the simplest case, the decay of an ensemble of systems
follows the well known law $N(t) = N_0$ $\exp[-\Gamma t]$. A
collection of solar system should follow such a law whenever the
probability of decay (planetary ejection or accretion) is constant in
time.  As shown in Fig. 1, the four ensembles of this paper come
close to following an exponential law. Nonetheless, the numerical
results depart from purely exponential behavior when the number $N(t)$
of surviving systems is small.

This trend can be quantified. If we assume that the basic decay 
law has an exponential form, then at any given time each solar 
system has a probability $q = {\rm e}^{-\Gamma t}$ of surviving 
and a probability $p=1-q$ of having already decayed. The distribution 
of surviving (undecayed) systems thus has a binomial form
\be
P(n) = \sum_{n=0}^N C_{Nn} q^n p^{N-n} \, , 
\ee
where $N$ is the total number of systems (100) and $n$ is the number
of solar systems surviving to time $t$. From this distribution, we 
can calculate the expectation value $\langle n \rangle$ = $q N$, the 
second moment $\langle n^2 \rangle$ = $q^2 N (N-1) + q N$, and the
width of the distribution $\sigma$ = $[q (1-q) N]^{1/2}$. As expected,
the relative width of the distribution grows larger as the number
$N(t)$ of surviving systems decreases. In particular, a measure of 
the relative width can be written in the form 
\be
{\sigma \over \langle n \rangle} = 
\Biggl( {1-q \over qN} \Biggr)^{1/2} \, = \, 
N^{-1/2} \bigl[ {\rm e}^{\Gamma t} - 1 \bigr]^{1/2} \, . 
\label{eq:error} 
\ee

To estimate the decay time scale, we fit a straight line to the
semi-log curves in Fig. 1. The fitting procedure is weighted by the
errors in the data and Eq. [\ref{eq:error}] is employed to estimate
the errors (due to the ever shrinking sample size).  The slope of the
fitted line determines the value of $\Gamma$, which in turn defines
the exponential decay time $\tau_{\rm decay}$ $\equiv$ $\Gamma^{-1}$.
The resulting time scales are listed in Table I for the four ensembles
of solar systems.  Because of the weighted fitting procedure, the
straight line fits agree much better with the numerical results at
shorter times; this behavior is due to the larger uncertainties that
arise for longer decay times (due to small $N$ statistics). For all
four planetary mass distributions, the formal uncertainties in the
half-lifes are somewhat less than 2 percent. The decay time scales, as
defined here, are relatively short, only 20,000 -- 125,000 years. The
half-life -- the time required for half of the systems in the ensemble
to decay -- is related to the decay time via $\tau_{1/2} = \tau_{\rm
decay} (\ln 2)$.

The temporal evolution of these systems can be understood in terms 
of a dynamical relaxation process. For a wide variety of assumptions, 
the dynamical relaxation time scale can be written in the general 
form 
\be
\tau_R = P_{\rm orb} Q \Bigl( {M_\star \over m_P} \Bigr)^2 \, , 
\label{eq:relax}
\ee
where $P_{\rm orb}$ is the orbital period for a representative orbit,
$M_\star$ is the mass of the central star, $m_P$ is the mass of the
planets, and $Q$ is a dimensionless factor that depends on the
geometry, density, and other characteristics of the system.  For
example, the factor $Q$ depends on whether the system is spherical or
nearly planar, and can also include a logarithmic correction factor
(e.g., see Binney and Tremaine 1987; Papaloizou and Terquem 2001).

For systems with equal mass planets, Eq. [\ref{eq:relax}]
predicts that the relaxation time scale should vary as the inverse
square of the planetary masses. Indeed, the ensemble with $m_P = 1
m_J$ has a decay time that is four times longer than that of the
ensemble with $m_P = 2 m_J$. The third ensemble, with a random
distribution of masses with mean mass 2 $m_J$, has a somewhat shorter
decay time.  This result also makes sense: The planets that are
ejected are generally the lower mass planets, which scatter off larger
ones. So even though the mean planet mass is 2 $m_J$, the mean mass of
the scatterers is higher and the time scale is correspondingly lower.

The decay time (or, equivalently, the half-life) defined above
describes how the initial ensemble of solar systems decays by removing
a planet. The resulting daughter products continue to eject or accrete
planets over longer time spans and we need to account for this
continued evolution.  To describe the longer term evolution, we
consider the entire ensemble of $N$ = 100 solar systems at once (for
each given mass function of planets).  This initial collection of 1000
planets is reduced over time, as calculated from the results of the
suite of numerical simulations.  By dividing the number of surviving
planets by $N$ = 100, we obtain an estimate of the ``typical''
evolution of a solar system over its first 1 Myr, i.e., an estimate of
the function $\np(t)$. The results are shown in Fig. 2 for the four 
ensembles of this paper.

The initial decay time $\tau_{\rm decay}$ is well defined because we
start the entire ensemble of solar systems in equivalent -- and known
-- initial states.  The longer term evolution of the planetary systems
is more complicated.  After a solar system decays by losing a planet,
the remaining daughter solar system will be more spread out (and hence
$Q$ will change in Eq. [\ref{eq:relax}]). To model this complication,
we consider the relaxation time to take the form $t_R \propto
\np^{-\beta}$, so that dynamical relaxation -- and hence planetary
ejection -- takes longer as the number of planets decreases (again,
see Binney and Tremaine 1987; Papaloizou and Terquem 2001). This form
for the relaxation time implies that the number of planets as a
function of time can be written in the form 
\be
\np(t) = {\np_0 \over \bigl[ 1 + \gamma t \bigr]^\alpha} \, , 
\label{eq:funevolve} 
\ee
where $\np_0$ = 10 (for these simulations), $\alpha = \beta^{-1}$, and
where $\gamma$ is determined by the magnitude of the relaxation
time. For each evolution function $\np(t)$ shown in Fig. 2, we
fit the results to a function of the form [\ref{eq:funevolve}].  The
resulting fits are shown as the solid curves in Fig. 2 and the
values of $\gamma$ and $\alpha$ are listed in Table I. Also listed in 
Table I is the evolution time scale $\tau_{\rm evolve}$, which is 
defined to be the time required (on average) for the number of planets 
in a solar system to be reduced by a factor of 2 (from $\np$ = 10 to 5).

Table I shows that the decay time $\tau_{\rm decay}$ or half-life
$\tau_{1/2}$ is closely related to the evolution time $\tau_{\rm
evolve}$. For each of the four ensembles of solar systems, the ratio
of time scales ${\cal R}$ = $\tau_{\rm evolve} / \tau_{1/2}$ has a
nearly constant value of $\cal R$ = 12 $\pm$ 1. In other words, for
these ensembles of 10 planet systems, the time required for a solar
system to lose half of its planets (the evolution time) is about 12
times longer than the time required for half of the population of
solar systems to decay by losing its first planet (the half-life).
Keep in mind that the half-life, as defined here, is the half-life 
of the initial 10 planet systems; the daughter products will contain 
fewer planets and have different (generally longer) half-lifes. 

\bigskip
\centerline{(Table I: Characteristic Time Scales)} 
\medskip

\subsection{Properties of the Resulting Solar Systems} 

Another result of this set of simulations is the distributions of
``final'' solar system properties. Since the first portion of this
study is limited to integrations of 1 Myr, the final solar system
properties are those at the 1 Myr mark. Solar systems can continue to
evolve over ever longer time frames -- even the planets in our own
(relatively sedate) solar system can change their orbital elements
over sufficiently long time periods (e.g., Laskar 1990; Duncan and
Quinn 1993). At the end of these 1 Myr integrations, the solar systems
have undergone substantial dynamical evolution from their initial
states and the suite of solar system properties is well defined.

For each simulation, we obtain the orbital elements of each surviving
planet at the end of one million years.  We find the distribution of
final orbital properties for these solar systems by considering the
entire collection of surviving planets and binning their orbital
elements. For our first ensemble of simulations, those with $m_P$ = 1
$m_J$, Fig. 3 shows the resulting distributions of system properties:
(a) the number of surviving planets, (b) semi-major axis, (c)
eccentricity, and (d) inclination angle measured with respect to the
original plane of the solar system. Figs. 4 -- 6 show the same sets of
distributions for the other three ensembles of simulations. Several
trends are evident from these distributions, as outlined below.

The number of surviving planets (after 1 Myr) shows a reasonably wide
distribution for all four ensembles. However, the case of $m_P$ = 1
$m_J$ planets shows the widest distribution as well as the largest
number of surviving planets. This result is due to two effects. First,
the time scales for dynamical relaxation are longer and hence this
ensemble of solar systems is dynamically less evolved. Second, the
lower mass of the planets allows for larger numbers of planets to
survive. To understand the importance of these two effects, we can
compare the simulations with $m_P$ = 1 $m_J$ at 1 Myr with the $m_P$ =
2 $m_J$ simulations at 0.25 Myr (these two classes of systems should
be dynamically relaxed to the same extent).  The evolutionary trends
shown in Figure 2 indicate that the $m_P$ = 1 $m_J$ solar systems
have, on average, 5.32 planets at the end of 1 Myr; for comparison,
the $m_P$ = 2 $m_J$ solar systems have 5.43 planets at 0.25 Myr. The
uncertainty in these mean values, due to the finite sample size
$(N=100)$ is about $\pm$ 0.1 planets, so that most of the difference
between the two sets of simulations can be attributed to the
difference in evolutionary time scales. (For further discussion of the
effects of time scales, see \S 3.6 and Table IV for the results of 10
Myr simulations.) 

For all four ensembles, the distribution of semi-major axis for the
surviving planets exhibits a well-defined peak at modest values ($a
\sim 10$ AU) and a long tail extending out to 150 AU (five times the
original outer edge of the solar systems). As expected, the detailed
shape of the distribution varies with the mass distribution of planets
(compare Figs. 3, 4, 5, and 6). For all four mass choices, the
distribution of eccentricity for the surviving planets spans the full
range $0 \le \epsilon \le 1$, with a broad peak near the center of the
range $\epsilon \sim 1/2$.  The distribution of inclination angles
(measured with respect to the starting plane of the system) shows a
somewhat broad, but still well-defined peak at $i = 20 - 30$ degrees. 
Although the distributions of $i$ extend up to $i=90$ degrees, they 
fall off rapidly from the peak near 30 degrees. 

The distributions of ``final'' system properties can be characterized
by finding the expectation value $\langle x \rangle$ and the standard
deviation $\sigma$ for each variable $x$. These results are presented
in Table II, where each entry is written in the form $\langle x \rangle
\pm \sigma$ (the variables are the number $\np$ of surviving planets,
semi-major axis $a$, eccentricity $\epsilon$, and inclination angle
$i$).  Keep in mind that the standard deviation $\sigma$ represents
the width of the distribution. The expectation values $\langle x
\rangle$ are determined to within an uncertainty given by $\delta
\langle x \rangle$ $\approx$ $\sigma/\sqrt{N}$ = $\sigma/10$.

The masses of the surviving planets represent another physical
variable to consider -- especially for the ensemble of solar systems
with randomly chosen masses. The surviving planets tend to have larger
masses than the ejected ones. The expectation value and standard
deviation for the mass distribution of surviving planets are also
included in Table II. If the surviving planets were drawn from the same
mass distribution as the starting state, the mean would be only 2.0
$m_J$.  The evolution from $\mbar$ = 2.0 $m_J$ to $\mbar$ = 2.9 $m_J$
is thus the signature of the mass segregation process acting in these
systems.

\bigskip
\centerline{(Table II: Solar System Properties at 1 Myr)} 
\medskip  

During the course of dynamical relaxation, the loss of planets can
take place through two different channels: outright ejection from the
solar system or through a collision with the central star (and the
subsequent accretion of the planet). The last line in Table II lists
the percentage of planets that are lost through collisions with the
central star for the four mass distributions. The other planets are
lost through ejection events. Notice that as the distribution of
planetary masses becomes wider, a smaller fraction of planets are lost
through collisions with the star (a larger fraction are ejected).

Although the solar systems are not fully relaxed at the end of the 1
Myr integration interval, most of the ejections yet to take place will
occur in the outer parts of the solar systems. The ensemble with $m_p$
= 1 $m_J$ has a longer relaxation time and hence is less evolved at
the 1 Myr mark. Comparison of its distributions (Fig. 3) with those
of the other mass distributions (Figs. 4 -- 6) show that mainly the
outermost planets will be ejected in the future. This result makes
sense because the outer parts of the system have a longer dynamical
time and are less evolved. By comparison, the inner parts of the
solar systems (say, where $a < 10$) are expected to evolve much less.

\subsection{Inner Planets and their Orbital Elements} 

For the class of solar systems studied here, the innermost surviving
planet is often in an orbit with $a \sim 1-2$ AU. These innermost
planets are the ones that would be detected in the ongoing planet
searches using radial velocity measurements. It is thus of interest to
characterize the orbital properties of these inner planets and compare
the results to those found in observational surveys.  The orbital
parameters for the innermost planets are summarized in Table III.

\bigskip
\centerline{(Table III: Orbital Properties of the Innermost Planets)} 
\medskip 

Distributions of the orbital parameters for the inner planets are
shown in Fig. 7. The upper panel shows the cumulative distribution of
semi-major axis $a$ for each of the four choices for the starting mass
distribution.  In all four cases, the distribution has the majority of
the planets inside 5 AU -- the starting inner boundary for the
simulations. The differential distribution (not shown) rises toward
smaller values of $a$ and reaches a well-defined peak near $a=1$ AU.
In contrast, the distribution of eccentricity shows that $\epsilon$ is
equally likely to be anywhere in the range $0 \le \epsilon \le 1$.
For comparison, the bottom panel of Fig. 7 also shows the cumulative
eccentricity distribution for observed extrasolar planets with $a \ge
0.1$ AU. The mass distribution with logarithmically spaced masses
comes closest to matching the observed distribution of eccentricities,
but all of the simulations produce too many planets at the highest
values of $\epsilon$.  Notice also that the inclination angle does not
matter for comparison with observed single-planet systems. If only the
innermost planet is detected, the plane of its orbit defines the plane
of the system, independent of its initial orbital plane.

The distributions of semi-major axes $a$ shown in Fig. 7 can be
characterized as follows.  For three of the four mass distributions,
approximately one third of the solar systems have an inner planet with
$a \le 2$ AU; the mass distribution with $m_P$ randomly distributed in
$\log m$ shows somewhat fewer planets with such small $a$.  For all
four mass distributions, however, nearly half of the systems produce
inner planets with $a \le 3$ AU. These results show that the inward
migration of giant planets (to $a \sim 1$ AU) through dynamical
relaxation in crowded solar systems is easily achieved. On the other
hand, this mechanism is not effective for moving planets much further
inward, unless additional giant planets are present at smaller radii. 
The simulations show a relatively well-defined boundary at $a$ = 1 AU,
beyond which no planets are found.

This effective boundary at 1 AU can be understood on energetic
grounds: Since all of the planets start at larger radii with $a \ge 5$
AU, migration to 1 AU results in the orbital binding energy increasing
by (at least) a factor of 5. In order to conserve energy, the ejected
planets (or those scattered to large $a$) must remove this energy from
the orbit of the migrating planet. Note that the ejected planets have
total energy close to zero, i.e., they carry away a relatively small
amount of energy. Simple accounting shows that for one planet to move
to 1 AU, the energy released is enough to disperse nearly all of the
other 9 planets in the system. This argument can be made more
rigorous: The Appendix presents an idealized calculation of the mass
and density profile for a disk of scattering bodies that is maximally
efficient at moving planets inward.

Although the solar systems are not necessarily fully relaxed as a
whole, the inner portions are relatively stable. As one measure of the
stability of the innermost planets, we can find the ratio of $a_1/a_2$
for each system, where $a_1$ ($a_2$) is the semi-major axis of the
first (second) planet. For the solar systems that result in an inner
planet with $a_1 \le 2$ AU (i.e., those cases that are candidates for
explaining the observed eccentric giant planets), the average value of
the ratio $\langle a_2/a_1 \rangle$ is near 20. For comparison, the
dynamical stability limit for a large companion with periastron $p_2$
interacting with a smaller planet with orbital radius $a_1$ is only
$p_2/a_1 \approx$ 7 (David et al. 2002; see also Gladman 1993). As a
second measure of stability, we can consider the ratio $R$ of the
periastron of the second planet to the apastron of the innermost
planet, i.e., $R$ = $a_2 (1 - \epsilon_2)/(a_1 (1 + \epsilon_1))$.
Fig. 8 plots the cumulative distribution of $R$ for the four sets of
simulations. For the uniform and logarithmic random mass
distributions, the ratio $R \ge 2$ for the vast majority of the solar
systems. For the two mass distributions with a single mass value,
however, a few of the solar systems have $R < 1$ (crossing orbits) and
$R \sim 1$ (unstable orbits). These solar systems will thus undergo
further dynamical evolution. Nonetheless, most of the solar systems
have relatively large values of $R$ (for all four mass distributions)
and are expected to be stable. 

This analysis shows that dynamical relaxation is efficient at
producing stable planetary orbits with semi-major axis $a \sim 1$ AU
and the full range of possible eccentricity. As a result, the
immediate output of these simulations can account for a fraction of
the orbits of observed extra-solar planets. Nonetheless, the
distributions of orbital properties differ from those of the observed
planets in two respects. The first is that the observed planets tend
not to have the very highest eccentricities. However, planets with
very high eccentricity tend to be the most unstable and are most
likely be affected by stellar tides (if the pericenter is not too much
larger than the stellar radius). As a result, they are least likely to
remain in the same orbits over longer times. The other discrepancy
(already mentioned) is that this mechanism -- by itself -- does not
produce many planets with shorter periods (significantly less than one
year). Keep in mind, however, that this result depends on the starting
conditions, where 10 planets are assumed to populate the range 5 AU
$\le a \le$ 30 AU.

\subsection{Chaotic Dynamics}  

A fundamental feature of crowded planetary systems is that they are
highly chaotic -- in the technical sense -- so that nearby
trajectories in phase space diverge exponentially in time (e.g.,
Ruelle 1989). For the dynamical systems studied in this paper, the six
phase space variables of all ten planets exhibit chaotic behavior. At
a given time, the systems can be characterized by a collection of
Lyapunov exponents, which represent the rate at which nearby
trajectories in phase space diverge. In this subsection, we obtain a
measure of these exponents $\lambda$ by using a shadowing method to
compute parallel trajectories for a collection of starting
configurations.

If the difference between nearby trajectories grows exponentially in
time, the growth rate $\lambda$ is defined through the relation 
\be 
\lambda (t) \equiv {\ln [ \Delta_X (t)/ 
\Delta_{X 0} ] \over t } \, , 
\ee
where $\Delta_X$ is the difference in a given phase space variable $X$
at time $t$ and $\Delta_{X 0}$ is the difference in that variable at 
the beginning of the time interval. For each trial, we obtain an
exponent $\lambda$ for each of the 10 planets and each of the 6 phase
space variables (60 values of $\lambda$). Although the exponents vary 
from case to case, the data show no discernible trends from planet to
planet or from variable to variable. By averaging the 60 exponents for 
a given trial, we obtain a well-defined mean value $\langle \lambda
\rangle$.  Through repeated trials, we estimate the mean exponent for
a longer time span. In practice, we sample the exponential divergence
every 1000 years, and integrate for a total of 30,000 years. 

For Jupiter mass planets, $\langle \lambda \rangle \approx 0.045$
yr$^{-1}$, which implies the corresponding time scale $\tau_\lambda =
\langle \lambda \rangle^{-1}$ $\approx$ 22 years. These planetary
systems are thus highly chaotic. With a growth time of only about 22
years, small perturbations (e.g., with starting amplitude $\sim
10^{-6}$) will become nonlinear in only 300 years. Since we are
interested in dynamical results on much longer time intervals, the
results of any simulation must be presented and interpreted
statistically.

\subsection{Scattering into Resonant Configurations} 

Within the collection of 11-body simulations that were performed, a
common outcome consists of two or three planets in a configuration
that exhibits orbital stability over time scales $\tau>10^{6}$ yr.  We
checked all such systems for low-order mean-motion commensurabilities
among the remaining planets, and found that $\sim 10$\% of the
configurations are {\it near} resonance after 1 Myr (roughly equally
distributed among the 2:1, 3:2, and 1:1 resonances).  In every case
examined, the libration widths of the resonant arguments are rather
large; this result is expected because the simulations lack a
dissipative mechanism (such as interactions with a remnant gaseous
disk). Over longer spans of time, these systems can evolve further and
many will leave resonance, due to the large libration widths and the
possible loss of additional planets from the system.  But resonant
configurations will survive in some cases and hence multiple-planet
scattering might account for systems such as HD 82943 (Udry et al.
2001) in which a pair of planets ($P_{b}=444.6 {\rm d}$, $P_{c}=221.6
{\rm d}$) have large eccentricities ($\epsilon_{b}=0.41$,
$\epsilon_{c}=0.54$) and experience large librations of the 2:1
resonant angles. A more extensive treatment should be done to
calculate the odds of systems attaining various resonant
configurations.

Several percent of the simulations yield systems in which two planets
participate in a 1:1 co-orbital resonance. An example is shown in
Fig. 9. The panel at the lower left shows the time history of the
system during the final million years of the simulation. The figure
also shows the quantities $a(1-\epsilon)$, and $a(1+\epsilon)$ for
each of three surviving planets in the configuration. The outer two
planets are participating in a co-orbital ``eccentric'' resonance of
the type recently described by Laughlin \& Chambers (2002).  The
planets exchange energy and angular momentum over a secular time scale
of roughly 50,000 years, and represent a stable (albeit complex)
resonant configuration. Sample trajectories of the planets are plotted
in the upper two frames of the figure, while the radial velocity curve
of the star (over 15 orbital periods) is shown at the lower right. In
this radial velocity curve, the stellar reflex velocity due to the
inner planet has been removed. The plotted residual thus represents
the influence of the dynamically interacting 1:1 resonant pair. This
residual radial velocity curve maintains periodicity, but it deviates
significantly from purely Keplerian motion, and shows complicated
variations even on timescales that are considerably shorter than the
$\sim 5 \times 10^{4}$ year period for secular eccentricity exchange.

\subsection{Additional Numerical Experiments} 

The primary ensembles of simulations (described above) provide a basic
understanding of the dynamics of these crowded planetary systems over
the first 1 Myr.  Nonetheless, the parameter space available for such
systems is extensive and several issues remain. One important issue is
to examine the longer term evolution of these systems. Another issue
is that the innermost planet in the simulations can easily migrate
inward to $a \sim 1$ AU (for starting configurations with $a \ge 5$
AU), but typically no further. Since observed eccentric giant planets
can reside at smaller values of $a$, it is useful to explore alternate
mass distributions to see if further inward migration can be attained.
In this subsection, we address these issues through additional sets of
numerical simulations.

To study the longer-term evolution of these systems, we have carried
out additional simulations using a longer time interval -- 10 million
years. We have performed these integrations for two of the mass
distributions: case [C], where the masses are selected randomly (but
uniformly) from the range 0 $\le m_P \le 4 m_J$, and case [D], where
the masses are chosen randomly in $\log m$ from the range $-1 \le
\log_{10} [m_P/m_J] \le 1$. In all cases, the simulations are started
with 10 planets in circular orbits that are logarithmically spaced
over the radial range 5 AU $\le r \le$ 30 AU.  For both mass
distributions, we have performed $N=50$ simulations. 

The resulting solar systems (at an evolutionary time of 10 Myr) have a
distribution of properties, as summarized in Table IV. The upper half
of the table shows the mean values and widths of the distributions for
the number of surviving planets $\np$, the semi-major axis $a$, the
eccentricity $\epsilon$, the inclination angle $i$, and the planet
mass $m_P$ (these data should be compared with those in Table II, for
the corresponding systems at 1 Myr).  The lower portion of the table
gives the fraction of surviving planets with $a < 2$ AU, as well as
the values of semi-major axis $a_1$, eccentricity $\epsilon_1$, and
mass $m_1$ for those inner planets (with $a < 2$).  The ratio of the
semi-major axis of the second planet to that of the inner planet is
also listed (compare with Table III). 

\bigskip 
\centerline{(Table IV: Solar System Properties at 10 Myr)} 
\medskip  

These longer term simulations show how much dynamical evolution takes
place in these solar systems between 1 Myr and 10 Myr. The results of
the 10 Myr simulations are much like the shorter ones. The mean number
of surviving planets decreases from $\np \sim 2.5$ to $\np \sim 2$
over this time interval. The mean eccentricity also decreases and
suggests that the high eccentricity planets are the ones lost from the
system. The mean mass of the surviving planets increases, as expected 
if the smaller planets are more likely to be ejected.  

We also performed another set of simulations that start with one large
planet and 19 smaller bodies. The large planet has the mass of Jupiter
and begins with a semi-major axis $a_0$ = 5 AU (much like the orbit of
Jupiter in our solar system). The 19 smaller bodies have randomly
chosen masses that are uniformly distributed over the range $0 \le m
\le 0.5 \, m_J$.  These smaller bodies begin in circular orbits with
radii evenly distributed in $\ln r$ over the range 3 AU $\le r \le$ 30
AU. This wider range of starting orbital radii was used so that the
region inside the `Jupiter' would be populated by smaller bodies. In
this case, we have carried out $N=100$ simulations to obtain a good
statistical description of the outcomes.

The results show that the large planet migrates inward to become the
innermost planet in 75 percent of the trials.  For those cases in
which the large planet becomes the innermost planet, the relevant
orbital parameters exhibit distributions characterized by $\langle a
\rangle$ = 2.08 AU $\pm$ 1.28 AU, and $\langle \epsilon \rangle$ =
0.36 $\pm$ 0.20.  This starting configuration is efficient at
transporting the large planet inward. Half of the simulations end with
the large planet at $a < 2$ AU, and 20 percent of the trials result in
$a < 1.5$ AU. However, only one trial places the large planet with
semi-major axis $a < 1$ AU. This effective boundary at about 1 AU can
be understood on energetic grounds (see \S 3.3 and the Appendix).

\section{PLANET INTERACTIONS WITH BACKGROUND DISK TORQUES} 

The dynamical relaxation calculations of the previous section
illustrate the difficulty in moving planets inward past 1 AU (for
planets starting with $a \ge 5$ AU).  Although a wide range of final
system properties can be realized through dynamical relaxation of 10
planet systems, the observed extra-solar planets often reside in
orbits with shorter periods (smaller semi-major axes $a$). One way to
achieve shorter periods is through tidal interactions of the planet
with the gaseous disk that gave it birth (e.g., Goldreich and Tremaine
1980; Lin, Bodenheimer, and Richardson 1996; Bryden et al. 2000),
although this mechanism does not generally produce high
eccentricities. Another way to achieve shorter periods is through
gravitational scattering interactions with a disk of planetessimals
(Murray et al. 1998). This latter mechanism requires a great deal of
mass in solid materials inside the orbit of the giant planet (see
Murray et al. 1998; see also the Appendix for a limiting case). In
this section, we explore the implications of combining dynamical
relaxation of two planets with the inward forcing driven by tidal
interactions with a background nebular disk (see also Kley 2000;
Murray, Paskowitz, and Holman 2002).

Specifically, we set up the following type of numerical experiments:
Two planets are placed on widely spaced orbits. The inner planet is
started with an orbital period of 1900 days, corresponding to a
semi-major axis of about 3 AU (a stellar mass of $M_\ast$ = 1.0
$M_\odot$ is used throughout).  The initial period of the outer planet
is larger by a factor of $\pi 2^{1/4} \approx 3.736\dots$ This period
ratio is an irrational number (more precisely, a 14 digit
approximation to an irrational number), so the planets are not started
in resonance.  As the outer planet loses orbital energy and angular
momentum through tidal torques, however, it moves inward and the two
planets can eventually be caught in resonances (see, e.g., Lee and
Peale 2002), typically the 3:1 resonance in this case.  The initial
eccentricities of both planets are drawn from a random distribution in
the range $0 < \epsilon < 0.05$.  The planet masses are drawn
independently from a random distribution in the range $0 < m_P < 5
m_J$.  These numerical experiments are thus quite similar to those
carried out by Snellgrove, Papaloizou and Nelson (2001), in their
study of the GJ 876 system, and by Laughlin, Chambers \& Fischer
(2002) in their analysis of the 47 UMa system. Our goal here is to
build on these previous studies by producing a statistical
generalization of the generic migration problem with two planets and
an exterior disk -- a situation that we expect is quite common during
the planet formation process.

The outer planet in the system is assumed to be tidally influenced by
a background circumstellar disk. Instead of modeling the planet/disk
interaction in detail, however, we introduce a frictional damping term
into the dynamics. This damping force has the form ${\bf f}= -{\bf v}
\taudamp^{-1}$ and is applied to the outer planet at each time step,
so the outer planet is gradually driven inward. This damping force is
proportional to the velocity and defines a damping time scale
$\taudamp$. In this set of simulations, we set the damping time scale
to be $\taudamp = 3 \times 10^5$ yr. If disk accretion occurs through
viscous diffusion and can be modeled using an `$\alpha$-prescription',
we can find the value of $\alpha$ for this choice of time scale: The
disk accretion time $\taudisk = \varpi^2/\nu$, where the viscosity
$\nu = (2/3) \alpha v_T^2 \Omega^{-1}$ (see Shu 1992).  Writing the
disk scale height $H$ in the form $H$ = $v_T/\Omega$ ($v_T$ is the
sound speed), the accretion time becomes $\taudisk = 1.5 (\varpi/H)^2
\Omega^{-1} \alpha^{-1}$. If we evaluate the disk scale height $H$ and
rotation rate $\Omega$ for a temperature of $T = 70$ K at $\varpi$ = 7
AU (where the outer planet begins), we find $\alpha$ = $7 \times
10^{-4}$ for our adopted time scale $\taudamp$ = 0.3 Myr. This value
falls within the expected range $10^{-4} \le \alpha \le 10^{-2}$ (see
Shu 1992). Recent estimates of this damping time scale (e.g., Tanaka,
Takeuchi, and Ward 2002) are in basic agreement with the value chosen
here. In this treatment, the energy dissipation time scale is assumed 
to be independent of the planet's orbital eccentricity, although more 
complicated behavior is possible (Tanaka et al. 2002; Papaloizou and 
Larwood 2000).  

These simulations include three additional effects. The first is the
damping of the eccentricity of the outer planet by the circumstellar
disk. The same angular momentum exchange between the disk and the
planet that leads to orbital migration can also modify the eccentricity 
of the orbit (e.g., Snellgrove et al. 2001; Agnor and Ward 2002).  To
incorporate this effect, we allow the orbital eccentricity of the
outer planet to be damped on a time scale $\tau_{\rm ed}$, which we
consider as a free parameter. A wide range of effective values for
$\tau_{\rm ed}$ are possible (Snellgrove et al. 2001), but we adopt
the range of values $\tau_{\rm ed}$ = 1 -- 3 Myr for this study.

Next, we include relativistic corrections to the force equations
(e.g., Landau and Lifshitz 1975; Weinberg 1972). This force correction
drives the periastron of both planetary orbits to precess (in the
forward direction). The effect is greater close to the star, so the
inner planet experiences a greater precession, which acts to move the
two planets away from a resonant condition. For planets that orbit
sufficiently close to their stars, this precession can be effective in
keeping the planets out of a perfect resonance. Since resonant
conditions lead to more extreme growth of orbital eccentricity (which
drives the system toward instability), relativistic precession acts to
make planetary systems more stable. In these simulations, the planets
only rarely wander close enough to the star to make this effect
important, but it is nonetheless included.

Finally, for completeness, the simulations take into account energy
loss due to tidal interactions between the planets and the central
star. Since the planets in these simulations spend most of their time
relatively far from the star, where tidal interactions are negligible,
we adopt an approximate treatment of this effect.  Specifically, we
adopt the approximations advocated in Papaloizou and Terquem (2001),
where the force exerted on the planet due to tidal interactions can be
written in the form 
\be 
{\bf F} = - {G m_P R_\ast^5 \over C j r^{11} } \bigl[ 
r^2 {\bf v} - ({\bf r} \cdot {\bf v}) {\bf r} \bigr] 
{0.6 R_p^3 \over 1 + (R_p / R_\ast)^3 } \, , 
\ee
where $R_\ast$ is the stellar radius, $R_p$ is the distance of closest
approach for a {\it parabolic} orbit with angular momentum $j$, and 
$C = 2 \sqrt{\pi}/3$ is a dimensionless constant of order unity (see
Papaloizou and Terquem 2001 for further discussion). This approximation 
assumes that most of the influence occurs near the point of closest 
approach and that the time between encounters is long compared to 
the time for tidal interaction itself. This approximation is valid 
when the planetary orbit has high eccentricity, which is the case 
for the planets in these simulations (see also Press and Teukolsky 
1977). 

With the starting conditions described above, the numerical
experiments are integrated forward in time until only one planet
remains, or the integration time reaches one million years. The
general evolutionary trend can be described as follows. The planets
are started out of resonance, but the outer planet is forced inward by
the dissipative term (which represents the action of a circumstellar
disk) until the planets enter into a mean motion resonance, usually
the 3:1 resonance. The two planets then migrate inwards together, near
resonance, but the planetary interactions tend to increase the orbital
eccentricity of both bodies. The large eccentricities drive the planet
to exhibit ever-larger departures from the resonant condition. The
eccentricities increase until the system becomes unstable, and a wide
range of final system properties can result. 

In practice, we continue the simulations until one of the planets is
ejected or driven into the central star, or the two planets collide
with each other. The effective radius for collisions is taken to be
about 2 $r_J$, where we assume that the planets have not fully
contracted.  After a planet is lost (via ejection, accretion, or
collision), the simulation is stopped and the orbital elements of the
surviving planet are calculated. However, the orbital elements of the
surviving planet can continue to evolve (after a planet is lost) as
long as the disk is still present.  To account for this additional
evolution, we assume that the inner disk has a lifetime $\tau_{\rm
disk}$ randomly drawn from a uniform distribution (with $\tau_{\rm
disk} \le 1$ Myr). After the main integration is stopped, the orbital
elements of the surviving planet are corrected for energy dissipation
and eccentricity damping over the time for which the disk remains
intact.

The orbital elements of the remaining planets show a distribution of
properties, as summarized in Table V and Figure 10.  For both choices
of the eccentricity damping time scale, the numerical experiments end
with about 60 percent ejections, 20 percent accretion events, and 1
percent collisions. The remaining cases reach the stopping time of 1
Myr without losing a planet. As expected, the ejected planets are more
likely to be those that start as the outer planet (only about one
third of the ejection events remove the inner planet). The accreted
planets are almost exclusively the inner planets (in all but one
case). The average time for the first planet to be ejected -- for all
outcomes -- is about 0.5 Myr, roughly comparable to the viscous
damping time of $\taudamp$ = 0.3 Myr. Accretion events take the
longest, with an average time of 0.55 Myr; ejection events have a mean
time of 0.22 Myr; collisions take place the fastest with a mean time
of only 0.90 Myr.  For the case of accretion or ejection events (of
either planet), the distributions of semi-major axis $a$, eccentricity
$\epsilon$, and mass $m_P$ are similar. The collisions result in
significantly different orbital properties, with smaller eccentricity
$\epsilon$ and larger mass $m_P$. The other general trend that emerges
from this suite of simulations is that the systems that remain stable
over the entire 1 Myr integration time are those with the smallest
planets, with a mean mass of only 1.5 $m_J$ (compared to a mean mass
$m_P$ = 2.8 $m_J$ for the whole ensemble). 

\bigskip 
\centerline{(Table V: Surviving Planet Properties)} 
\medskip 

The distributions of semi-major axis and eccentricity for the
surviving planets are shown in Fig. 10, which includes results for
both choices of eccentricity damping time scale ($\tau_{\rm ed}$ = 1
Myr and 3 Myr). The distributions are shown both at the time when the
first planet is ejected and after additional orbital evolution has
taken place. This migration mechanism naturally populates the inner
region of the solar systems, i.e., the range of semi-major axies 0.1
AU $\le a \le$ 1 AU where dynamical relaxation (\S 3) is ineffective
(for planets that start with $a > 5$ AU).  The distribution of
semi-major axes is roughly consistent with that of the observed
population of extra-solar planets. However, the simulations tend to
produce too many planets with highly eccentric orbits compared to the
observed sample (see the bottom planel of Fig. 10).

Another way to compare the theoretical simulations with the observed
sample of extrasolar planets is through the $a-\epsilon$ plane.
Fig. 11 shows this plane for the observed planets and the two choices
of eccentricity damping time scale using in the numerical simulations.
Although the separate distributions of $a$ and $\epsilon$ show
reasonable agreement between the theory and the observed sample, the
two-dimensional distributions (in $a-\epsilon$ space) provide a
stronger test.  Fig. 11 shows that the observed planet population
contains more orbits with both small $a$ and small $\epsilon$ than the
theoretical model (the lower left region of Fig. 11), or, equivalently, 
the theory produces too many orbits with high eccentricity.  This
trend can be quantified using a two-dimensional Kolmogorov-Smirnov
(K-S) test on two samples. The K-S test provides the probability that
the two samples were drawn from the same underlying population. In
this case, the probability is rather low, less than one percent, and
hence the two samples are indeed different. Notice that this
discrepancy could be relieved by tidal circularization that takes 
place over time spans much longer than the 1 Myr time scale of these 
simulations; this mechanism acts to decrease the eccentricity of the
planetary orbits.

\section{SUMMARY AND DISCUSSION} 

This paper has explored the dynamical relaxation of giant planet
systems in their early phases of evolution. By integrating a large
number of equivalent realizations (e.g., $N=100$) for each set of
starting conditions, we obtain the distributions of the final system
properties. The motivation for this work is both to understand the
dynamics of these crowded planetary systems and to determine whether
or not dynamical evolution can produce planetary orbits like those
observed in the current sample of extra-solar planets.

The first dynamical result is a determination of the decay time for
each class of solar system. In this context, the decay time is the
time required for a solar system to either eject its first planet or
accrete a planet onto the central star. Any given sample of solar
systems has a well-defined half-life -- the time over which half of
the sample will decay (see Fig. 1 and Table I). Over longer spans of
time, the solar systems continue to evolve by spreading out, ejecting
planets, and accreting planets. The evolution of these systems can be
described by a simple function (see Eq. [\ref{eq:funevolve}] and
Fig. 2).  A related time scale is time required for a given solar
system to lose half of its initial population of planets. For the
ensembles studied here, this evolution time is about 12 times longer
than the corresponding half-life (Table I).

All of the solar systems studied here are highly chaotic in the
technical sense. To quantify this behavior, we have calculated
characteristic exponents for the orbital parameters in these
systems. The corresponding Lyapunov times -- roughly the e-folding
time for small departures between equivalent physical systems to grow
-- is relatively short, typically a few decades (10s of years).
Putting all of the timing results together, we obtain the following
ordering of time scales 
\be
\tau_{\rm orbit} \ll \tau_{\rm chaos} \ll 
\tau_{\rm decay} \ll \tau_{\rm evolve} \ll \tau_{\rm lifetime} \, , 
\label{eq:torder} 
\ee
where $\tau_{\rm lifetime}$ is the total expected lifetime of the 
systems (typically billions of years). 

The main astronomical motivation for this work is to account for the
observed extra-solar planets in highly eccentric orbits. For crowded
planetary systems initially populated in the radial range 5 AU $\le a
\le$ 30 AU, dynamical relaxation naturally produces eccentric orbits
with semi-major axis $a \sim 1$ AU. Such orbits occur readily; for
example, one third of the systems successfully place planets in orbits
with $a < 2$ AU (see Table III and Fig. 7). Although successful in
producing eccentric orbits with $a \approx 1$ AU, dynamical
relaxation, acting in isolation, does not generally drive planets to
migrate further inwards. The basic reason for this difficulty is that
these giant planet systems do not have scattering bodies at small
radii to remove further energy from the system.

The Appendix quantifies this difficulty and defines a benchmark disk
model which has the minimum mass necessary to cause planetary
migration through scattering events (subject to the idealizations of
the calculation). To move a planet of mass $m_P$ from a starting
semi-major axis $a_0$ inward to $a_f$, the required mass is $M_{MES} =
m_P \ln[a_0/a_f]$, and it must be located within the annulus $a_f \le
r \le a_0$ (provided that the mass moves on orbits of low
eccentricity).  For a giant planet to migrate inward beyond the
effective boundary at $\sim1$ AU found in the simulations, the disk
must contain at least a planetary mass worth of scattering bodies
close to the star ($a < 1$ AU). Additional migration by giant planet
scattering is thus problematic because giant planets do not readily
form (presumably) in the inner solar system (hence the need for
migration).  But additional inward migration by planetessimals --
another leading candidate -- is also problematic: Although the inner
solar system naturally produces such entities, they are made of heavy
elements. To move a Jupiter-mass planet from $a$ = 1 AU to $a = 1/3$
AU, for example, the disk must contain about 370 Earth masses of rocky
material {\it within 1 AU of the star}.

This work also demonstrates that dynamical instability in crowded
planetary systems can result in a pair of surviving planets which
are participating in large-amplitude librations around low-order
mean-motion resonances. In particular, multiple-planet scattering can
lead to pairs of planets in unusual co-orbital resonances, including
the 1:1 eccentric resonance shown in Fig. 9. Short-term dynamical
interactions of such a configuration would lead to a readily
detectable radial velocity signature.

As an alternate migration scenario, we have explored the possibility
of multiple giant planets being driven inward through the action of
tidal torques in a circumstellar disk. In this case, the outer planet
interacts with the disk and has energy and angular momentum drained
away from its orbit. As the outer planet migrates inward, it
eventually becomes close enough to the interior planet to drive
eccentricity growth and increasingly violent interactions. Such
systems are not stable in the long term and adjust themselves to
stability by ejecting a planet, accreting a planet onto the central
star, or by having the two planets collide. The surviving planet is
left on an eccentric orbit of varying semi-major axis, i.e., 0.1 $\le
a \le$ 3 AU (see Fig. 10 and Table V).
 
In addition to the specific results described above, the results of
this investigation illustrate a more general aspect of solar system
formation and dynamics. Although the planet formation process can
proceed through many different channels -- or at least many scenarios
for planet formation remain viable -- all of them lead to dynamical
systems that are highly chaotic (see Eq. [\ref{eq:torder}]).
Even the most sedate end result -- a well-ordered solar system like
our own -- displays chaotic behavior over sufficiently long spans of
time. In the face of such chaos, the results of the planet formation
process must be described in terms of a full {\it distribution} of
results. Given the enormous variation possible, and the extreme
sensitivity to initial conditions, it does not make sense to talk
about a single outcome of any given dynamical experiment; and this
result applies to both theoretical calculations of planet formation
and the experiments done by planet-forming systems in our galaxy.

\bigskip 
\centerline{\bf Acknowledgements} 
\medskip 

During the course of this work, we have benefited from discussions
with Gus Evrard and Elisa Quintana. We would also like to thank the
referees (Eric Ford and a second anonymous referee) for many useful
suggestions that improved this paper. This work was supported in part
by NASA through the Origins of the Solar System Program and by the
University of Michigan through the Michigan Center for Theoretical
Physics.

\newpage 
\centerline{\bf APPENDIX: THE MAXIMALLY EFFICIENT SCATTERING DISK} 
\medskip 

In this Appendix, we derive a limiting form for the surface density
distribution and disk mass necessary to drive planetary migration
through scattering interactions.  For the sake of definiteness, we
consider a migrating giant planet of mass $m_P$ and initial semi-major
axis $a_0$. The starting energy of its orbit is thus $E_0$ = $- G
M_\ast m_P / 2 a_0$. In order to migrate inward, the planet must have
energy removed from its orbit so that it falls deeper into the
gravitational potential well of its parental star. In this idealized
calculation, we assume that the requisite loss of energy takes place
through scattering interactions, where the scattering bodies have mass
$\mu$ and move on orbits of low eccentricity. The scattering bodies
could be either planetessimals (small $\mu$) or other planets (larger
$\mu$).  In this first approximation, each scattering interaction
takes place at radius $r$ and removes energy $\Delta E$ = $-G M_\ast
\mu /2r$ from the orbit of the planet. This energy increment is that
required for the scattering body to become unbound (at zero energy)
from a circular orbit at radius $r$.

After one scattering interaction, the semi-major axis of 
the migrating giant planet becomes 
$$
a_1 = a_0 (1 + \mu / m_P)^{-1} \, , \eqno({\rm A}1) 
$$
where we have assumed that the planet moves as far inward 
as possible without violating conservation of energy. The 
disk is assumed to be maximally efficient in the sense that 
for every new orbit the giant planet obtains, a new scattering 
body will be present to scatter and remove further energy 
from the planet's orbit. If the planet moves as far inward 
as possible, each scattering will decrease the semi-major 
axis by the same factor. After $n$ scattering interactions, 
the new orbit attain a semi-major axis $a_n$ given by 
$$
a_n = a_0 (1 + \mu / m_P)^{-n} \, . \eqno({\rm A}2) 
$$ 
Let $f_n \equiv a_0/a_n$ be the factor by which the semi-major axis
decreases after $n$ steps. The total mass $M_S$ in scattering bodies
is related to the mass of the bodies via $M_S = n \mu$.  The factor
$f_n$ can then be written in the form 
$$
f_n = \bigl( 1 + {M_S / m_P \over n} \bigr)^n \, . \eqno({\rm A}3) 
$$

For a fixed mass $M_S$ in scattering bodies, the factor $f_n$ increases
(slowly) with increasing $n$, so the maximum migration factor can be
defined by taking the limit $n \to \infty$. The maximum factor is thus
given by 
$$
\lim_{n \to \infty} f_n = {\rm e}^{M_S/m_P} \, . \eqno({\rm A}4) 
$$
In other words, the minimum mass $M_{MES}$ in scattering bodies 
required to move a planet from $a_0$ to $a_f$ can be written in 
the form 
$$
M_{MES} = m_P \ln [a_0 / a_f] \, . \eqno({\rm A}5) 
$$ 
For typical migration patterns, $a_0$ $\approx$ 5 AU and $a_f$
$\approx$ 0.1 AU, so $M_S > \ln[50] m_P$. If $m_P$ = $m_J$, for
example, then the mass $M_S$ of scattering bodies must exceed about 
4 Jupiter masses or 1200 times the mass of Earth. 

This result suggests that migration over large distances via
scattering processes is problematic for any likely mass of the
scattering bodies. If the scattering entities are giant planets, then
the mass requirements are reasonable, but the scattering bodies --
which are giant planets themselves -- must already reside at inner
locations in the solar system. The scattering calculations of this
paper show that giant planet systems will rarely scatter multiple
planets to small $a$ so that they can scatter off of each other to
even smaller $a$. Migration factors of $f=5$ are readily obtained,
whereas factors of $f=50$ are (almost) never realized in the numerical
experiments.

On the other hand, if the scattering bodies are planetessimals --
which are made of heavy elements -- then location is no longer a
problem, but the amount of mass required is excessive. For example,
our solar system contains a total of only 100 Earth masses worth of
rocky material, a factor of 13 too small to attain a migration factor
of $f$ = 50. If migration is limited to metal rich systems, and if
early solar systems have more mass than the minimum mass solar nebula,
then the total inventory of planetessimals could approach the
necessary minimum value. But even in such a favorable case, most of
the disk mass (and hence the rocky material) is expected to lie in the
outer solar system, whereas migration through scattering requires it
to reside in the inner solar system.

The maximally efficient scattering disk, as defined here, displays a
particular distribution of surface density. In discrete form, the
surface density can be written 
$$ 
\sigma (r) = \sum_{k=0}^{n_f - 1} {\mu \over 2 \pi r} 
\delta (r - a_k) \, , \eqno({\rm A}6) 
$$
where $\delta (x)$ is the Dirac delta function and the positions 
$a_k$ are determined through Eq. [A2]. Taking the limit 
$n \to \infty$, $\mu \to 0$, with $n \mu \to$ {\sl constant}, 
we obtain a surface density profile of the form 
$$
\sigma_{MES} (r) = {m_P \over 2 \pi r^2} \, . \eqno({\rm A}7) 
$$
As a consistency check, notice that if we integrate the surface density 
profile [A7] over an annulus of outer radius $r_0$ and inner radius 
$r_f$, the mass enclosed is given by $\Delta M$ = $m_P \ln [r_0/r_f]$, 
in agreement with Eq. [A5].

{} 

\newpage 

\bigskip
\centerline{\bf Table I: Characteristic Time Scales} 
\medskip 

\begin{center}
\begin{tabular}{lcccc}
\hline 
\hline
variable / sample & $m_P$ = 1 $m_J$ & $m_P$ = 2 $m_J$ & random $m_P$ & log-random\\
\hline 
$\tau_{\rm decay}$ (Myr) & 0.125 & 0.0345 & 0.0185 & 0.00721\\ 
$\tau_{1/2}$ (Myr) & 0.0866 & 0.0239 & 0.0128 & 0.0050\\ 
$\tau_{\rm evolve}$ (Myr) & 1.14 & 0.309 & 0.157 & 0.0569\\ 
\hline 
$\alpha$ & 0.932 & 0.596 & 0.440 & 0.304 \\ 
$\gamma$ (Myr)$^{-1}$ & 0.981 & 6.89 & 22.6 & 143 \\ 
\hline 
\hline 
\end{tabular}
\end{center} 

\vskip0.75truein 

\bigskip
\centerline{\bf Table II: Solar System Properties at 1 Myr} 
\medskip 

\begin{center}
\begin{tabular}{ccccc}
\hline 
\hline
variable / sample & $m_P$ = 1 $m_J$ & $m_P$ = 2 $m_J$ & random $m_P$ & log-random\\
\hline 
$\np$ & 5.2 $\pm$ 1.2 & 2.9 $\pm$ 0.79 & 2.5 $\pm$ 0.83 & 2.3 $\pm$ 0.82\\
$a$ (AU) & 49 $\pm$ 38 & 44 $\pm$ 43 & 33 $\pm$ 37 & 29 $\pm$ 33\\
$\epsilon$ & 0.53 $\pm$ 0.24 & 0.52 $\pm$ 0.22 & 0.42 $\pm$ 0.23 & 0.41 $\pm$ 0.20\\
$i$ (degrees) & 36 $\pm$ 24 & 34 $\pm$ 23 & 30 $\pm$ 21 & 21 $\pm$ 20\\ 
\hline 
$m_P$ ($m_J$) & 1 $\pm$ 0 & 2 $\pm$ 0 & 2.9 $\pm$ 0.86 & 4.7 $\pm$ 2.9\\ 
\hline
accretion events ($\%$) & 23 & 15 & 11 & 5.5 \\ 
\hline 
\hline 
\end{tabular}
\end{center} 

\newpage 

\bigskip
\centerline{\bf Table III: Orbital Properties of the Inner Planets at 1 Myr} 
\medskip 

\begin{center}
\begin{tabular}{ccccc}
\hline 
\hline
variable / sample & $m_P$ = 1 $m_J$ & $m_P$ = 2 $m_J$ & random $m_P$ & log-random\\
\hline 
${\cal F}(a<2{\rm AU})$ & 0.37 & 0.29 & 0.33 & 0.09 \\ 
$a$ (AU) & 1.4 $\pm$ 0.28 &  1.3 $\pm$ 0.28 & 1.6 $\pm$ 0.24 & 1.6 $\pm$ 0.37 \\
$\epsilon$ & 0.65 $\pm$ 0.16 & 0.56 $\pm$ 0.22 & 0.48 $\pm$ 0.21 & 0.61 $\pm$ 0.20 \\
$a(1-\epsilon)$ (AU) & 0.47 $\pm$ 0.23 & 0.59 $\pm$ 0.30 & 0.83 $\pm$ 0.33 & 0.60 $\pm$ 0.33\\ 
$R = a_2/a_1$ & 15 $\pm$ 7.5 & 20 $\pm$ 14 & 21 $\pm$ 18 & 15 $\pm$ 9.7 \\ 
\hline 
$m_P$ ($m_J$) & 1 $\pm$ 0 & 2 $\pm$ 0 & 3.0 $\pm$ 0.52 & 4.3 $\pm$ 2.4\\ 
\hline 
\hline 
\end{tabular}
\end{center} 

\vskip 0.75truein 

\bigskip
\centerline{\bf Table IV: Solar System Properties at 10 Myr} 
\medskip 

\begin{center}
\begin{tabular}{ccc}
\hline 
\hline
variable / sample & random $m_P$ & log-random\\
\hline 
$\np$ & 1.9 $\pm$ 0.54 & 2.0 $\pm$ 0.58\\ 
$a$ (AU) & 32 $\pm$ 43 & 24 $\pm$ 29 \\
$\epsilon$ & 0.49 $\pm$ 0.22 & 0.37 $\pm$ 0.21\\
$i$ (degrees) & 23 $\pm$ 19  & 18 $\pm$ 16 \\ 
$m_P$ ($m_J$) & 3.0 $\pm$ 0.70 & 5.0 $\pm$ 2.6 \\ 
\hline 
${\cal F}(a<2{\rm AU})$ & 0.36 & 0.060 \\ 
$a_1$ (AU) & 1.6 $\pm$ 0.27 & 1.4 $\pm$ 0.38 \\
$\epsilon_1$ & 0.58 $\pm$ 0.20 & 0.66 $\pm$ 0.07 \\ 
$m_1$ ($m_J$) & 3.1 $\pm$ 0.50 & 3.3 $\pm$ 0.43 \\
$R = a_2/a_1$ & 25 $\pm$ 24 & 11 $\pm$ 7.8\\ 
\hline 
\hline 
\end{tabular}
\end{center} 

\newpage 
\bigskip 
\centerline{\bf Table V: Surviving Planet Properties:} 
\centerline{\bf Two planets systems tidally driven by a disk (350 Trials)}  
\medskip 

\begin{center}
\begin{tabular}{lccccc}
\hline 
\hline
variable/sample & $\tau_{\rm ed}$ = 1 Myr & $\tau_{\rm ed}$ = 3 Myr \\
\hline 
$a$ (AU) & 0.82 $\pm$ 1.25 & 0.91 $\pm$ 0.95 \\
$\epsilon$ & 0.45 $\pm$ 0.27 & 0.53 $\pm$ 0.29 \\
$m_P$ ($m_J$) & 2.9 $\pm$ 1.4 & 2.7 $\pm$ 1.4 \\ 
$\tau_{\rm eject}$ (Myr) & 0.49 $\pm$ 0.39 & 0.53 $\pm$ 0.40\\  
Accretion fraction ($\%$) & 20 & 19 \\ 
Ejection fraction ($\%$) & 62 & 56 \\ 
Collision fraction ($\%$) & 0.5 & 1.2 \\ 
\hline 
\hline 
\end{tabular}
\end{center}

\newpage 
\begin{figure}
\figurenum{1}
\epsscale{1.0}
\plotone{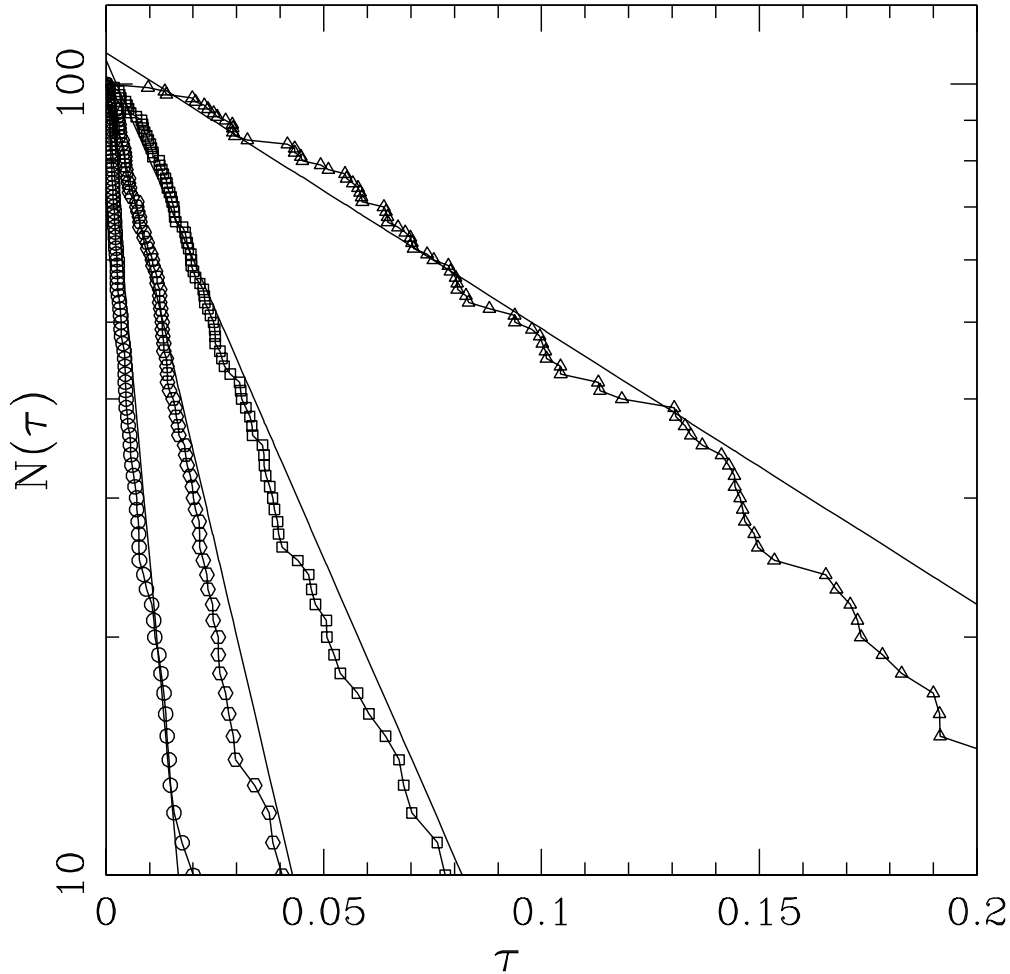}
\figcaption{Determination of the decay time for crowded solar systems.
For an ensemble of 100 numerical simulations, each curve depicts the
number $N(\tau)$ of solar systems that have not decayed by ejecting or
accreting a planet. The functions $N(\tau)$ are plotted versus time
(in millions of years). The triangles show the results of simulations
using planets with $m_P$ = 1 $m_J$; the squares show the results for
planets with $m_P$ = 2 $m_J$; the hexagons show the results for
planets with randomly chosen masses in the range $0 \le m_P \le 4
M_J$; the circles show the results for a planet mass distribution 
chosen uniformly in $\log m$ within the range $-1 \le \log_{10} 
[m_P/m_J] \le 1$. The solid lines are weighted fits to the numerical 
results (see text) and the slope of the lines determine the decay 
time (see Table I).}  
\end{figure} 

\newpage 
\begin{figure}
\figurenum{2}
\epsscale{1.0}
\plotone{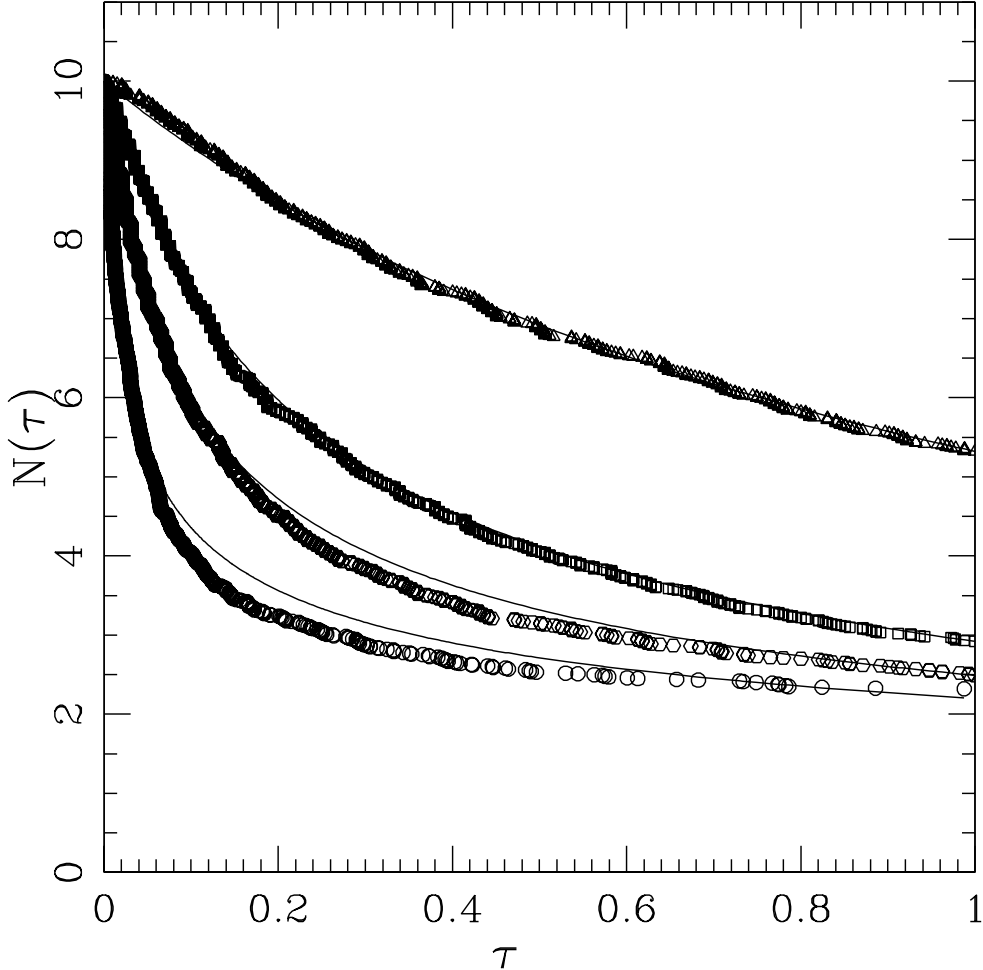}
\figcaption{Determination of the evolutionary behavior for crowded
solar systems. For each starting mass function, the entire ensemble of
$N=100$ simulations collectively determines the typical behavior for
the number of surviving planets as a function of time (in millions of
years). The triangles show the results for simulations using planets
with $m_P$ = 1 $m_J$; the squares show the results for planets with
$m_P$ = 2 $m_J$; the hexagons show the results for planets with
randomly chosen masses in the range $0 \le m_P \le 4 M_J$; the circles 
show the results for planets with masses chosen uniformly in $\log m$ 
in the range $-1 \le \log_{10} [m_P/m_J] \le 1$. The solid curves are 
fitted functions of the form given by Eq. [\ref{eq:funevolve}] 
(see text).} 
\end{figure} 

\newpage 
\begin{figure}
\figurenum{3}
\epsscale{1.0}
\plotone{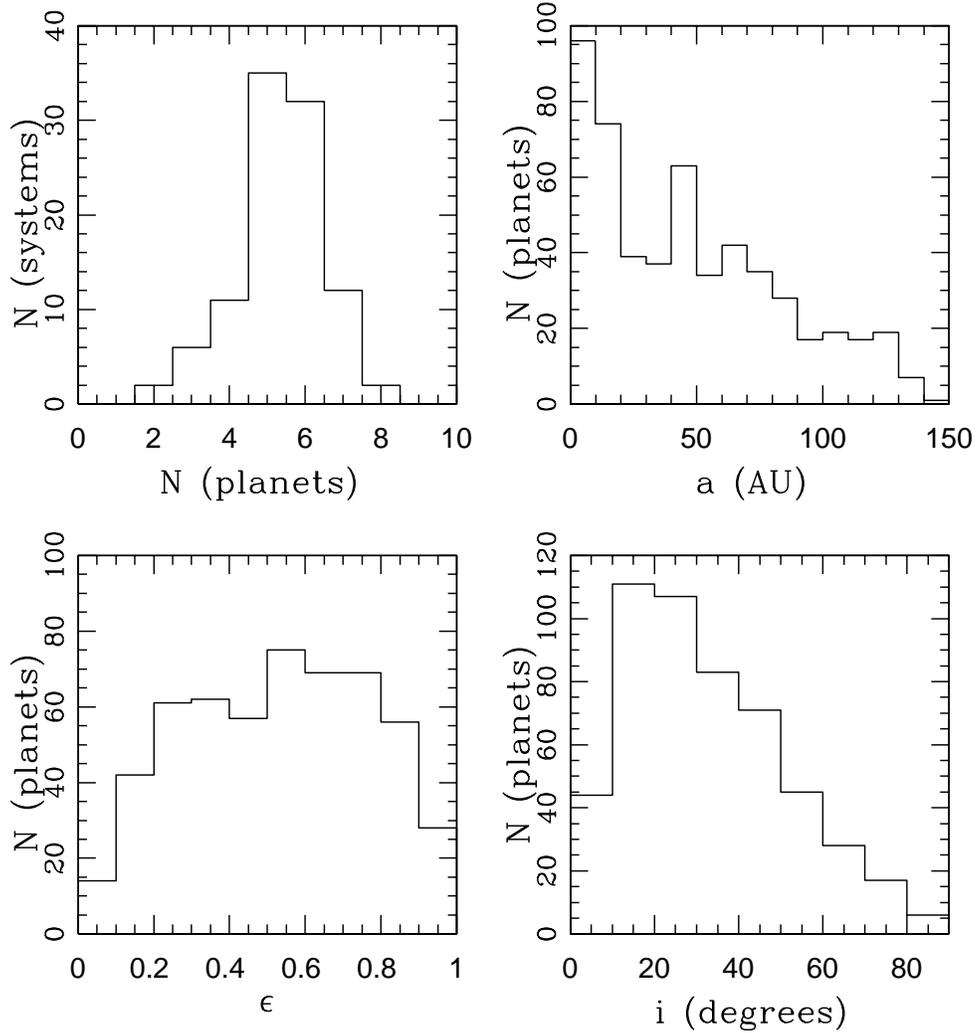}
\figcaption{Characterization of solar system properties after 1 Myr of
dynamical evolution -- for the case of planets with $m_P$ = 1 $m_J$.
All of the remaining planets in the ensemble of $N=100$ solar systems
are folded together in these histograms. 
(a) Distribution of number of remaining planets. 
(b) Distribution of semi-major axes of surviving planets.
(c) Distribution of eccentricity of surviving planets.
(d) Distribution of inclination angle of surviving planets. }  
\end{figure} 

\newpage 
\begin{figure}
\figurenum{4}
\epsscale{1.0}
\plotone{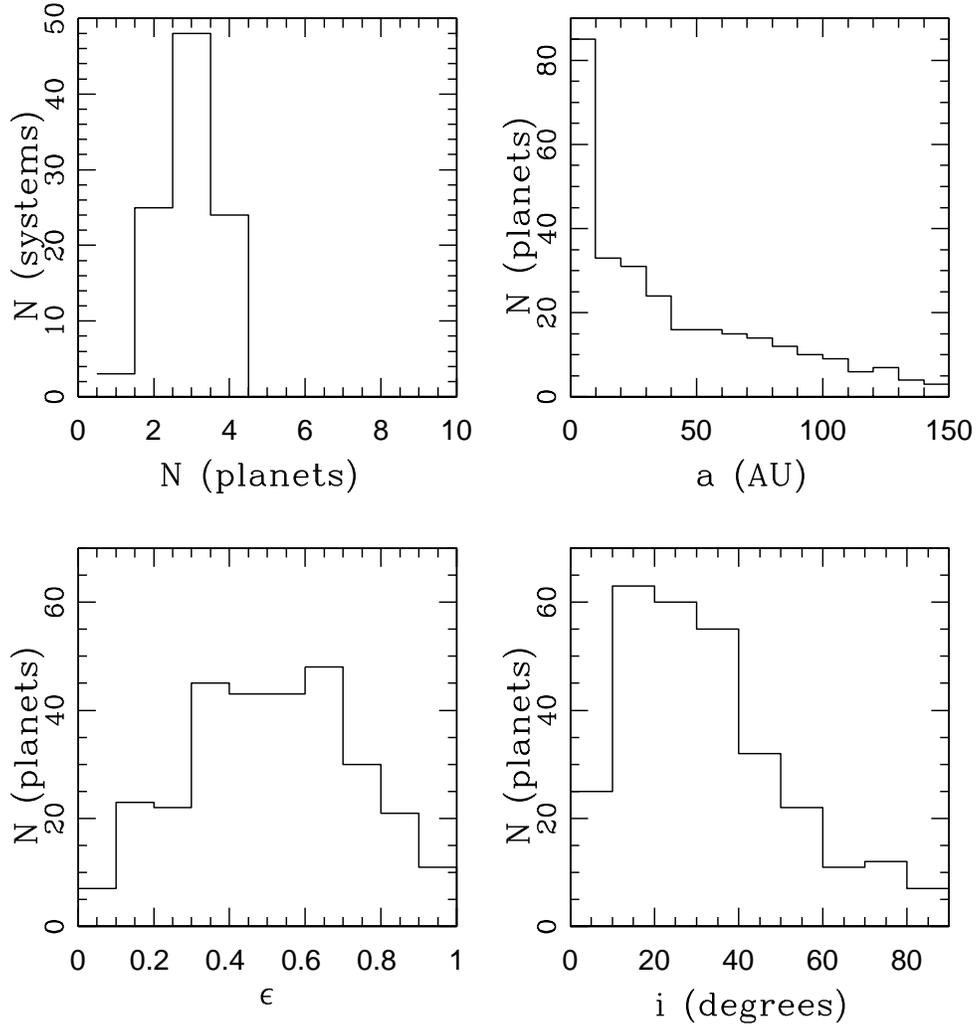}
\figcaption{Characterization of solar system properties after 1 Myr of
dynamical evolution -- for planets with $m_P$ = 2 $m_J$. The remaining
planets from all $N=100$ solar systems are folded together in these
histograms.  
(a) Distribution of number of remaining planets.  
(b) Distribution of semi-major axes of surviving planets.  
(c) Distribution of eccentricity of surviving planets.  
(d) Distribution of inclination angle of surviving planets. }  
\end{figure} 

\newpage 
\begin{figure}
\figurenum{5}
\epsscale{1.0}
\plotone{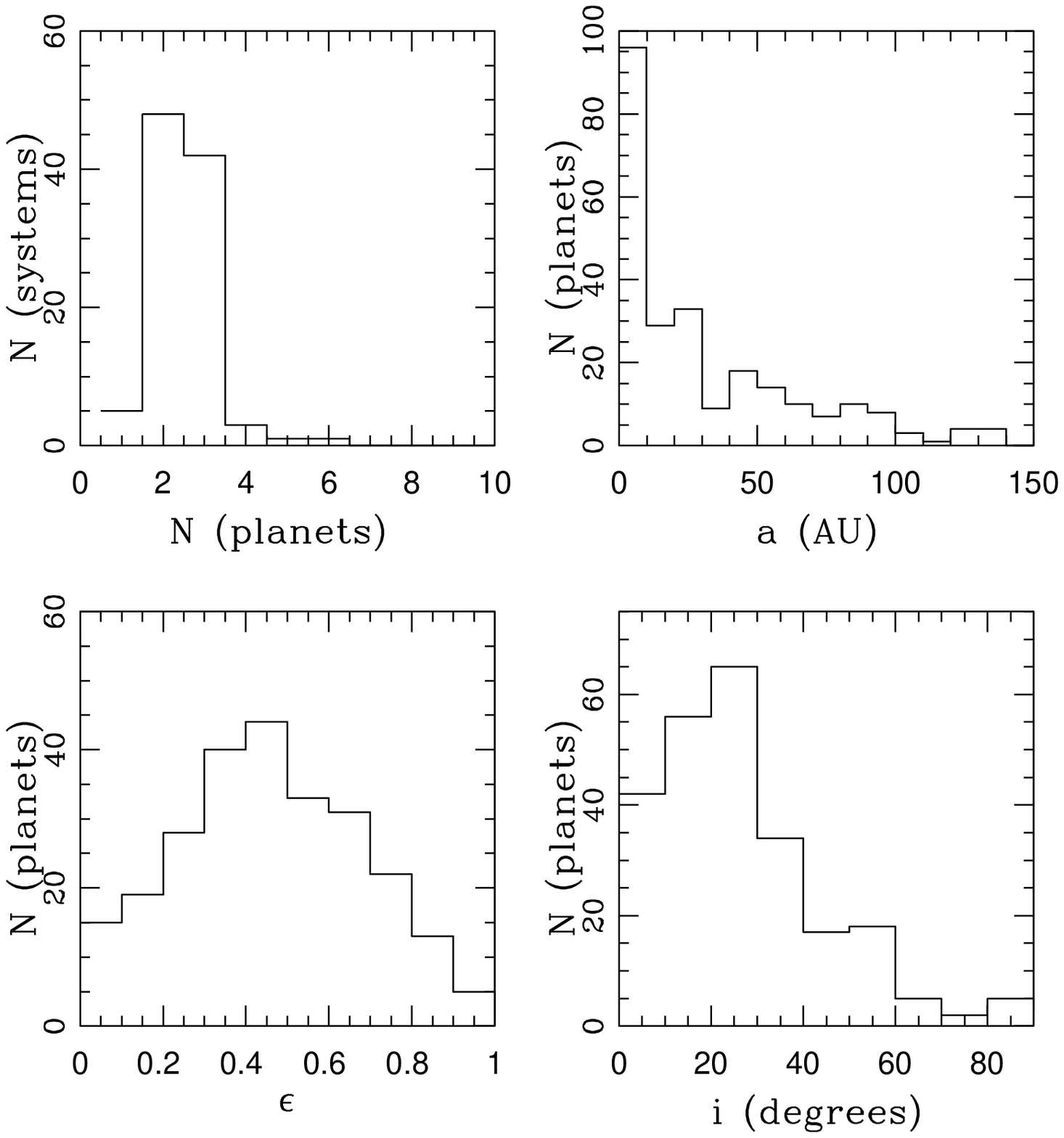}
\figcaption{Characterization of solar system properties after 1 Myr of
dynamical evolution -- for planets with randomly chosen masses within
the range $0 \le m_P \le 4 m_J$.  The remaining planets from all 
$N=100$ solar systems are folded together in these histograms. 
(a) Distribution of number of remaining planets. 
(b) Distribution of semi-major axes of surviving planets.
(c) Distribution of eccentricity of surviving planets.
(d) Distribution of inclination angle of surviving planets. }  
\end{figure} 

\newpage 
\begin{figure}
\figurenum{6}
\epsscale{1.0}
\plotone{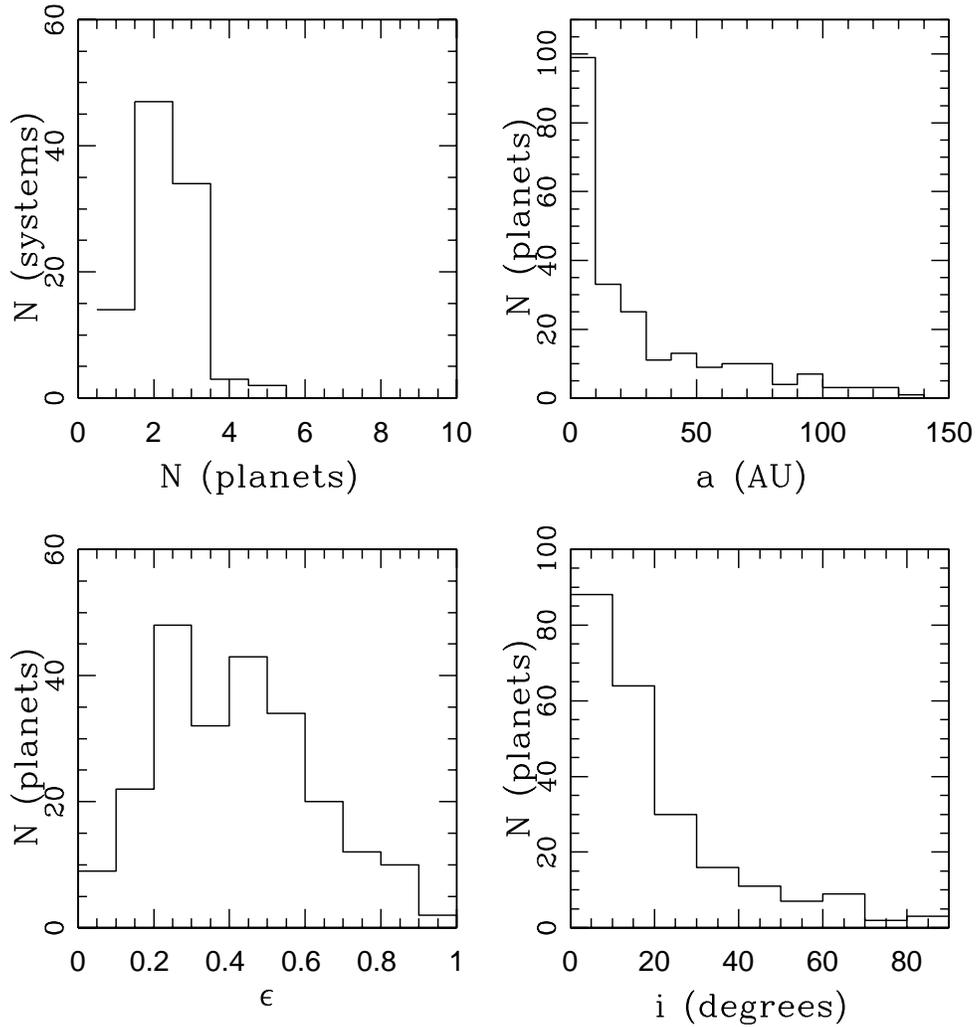}
\figcaption{Characterization of solar system properties after 1 Myr 
of dynamical evolution -- for planets with masses chosen randomly in 
$\log m$ within the range $-1 \le \log_{10} [m_P/m_J] \le 1$.  
The remaining planets from all $N=100$ solar systems are folded 
together in these histograms. 
(a) Distribution of number of remaining planets. 
(b) Distribution of semi-major axes of surviving planets.
(c) Distribution of eccentricity of surviving planets.
(d) Distribution of inclination angle of surviving planets. }  
\end{figure} 

\newpage 
\begin{figure}
\figurenum{7}
\epsscale{1.0}
\plotone{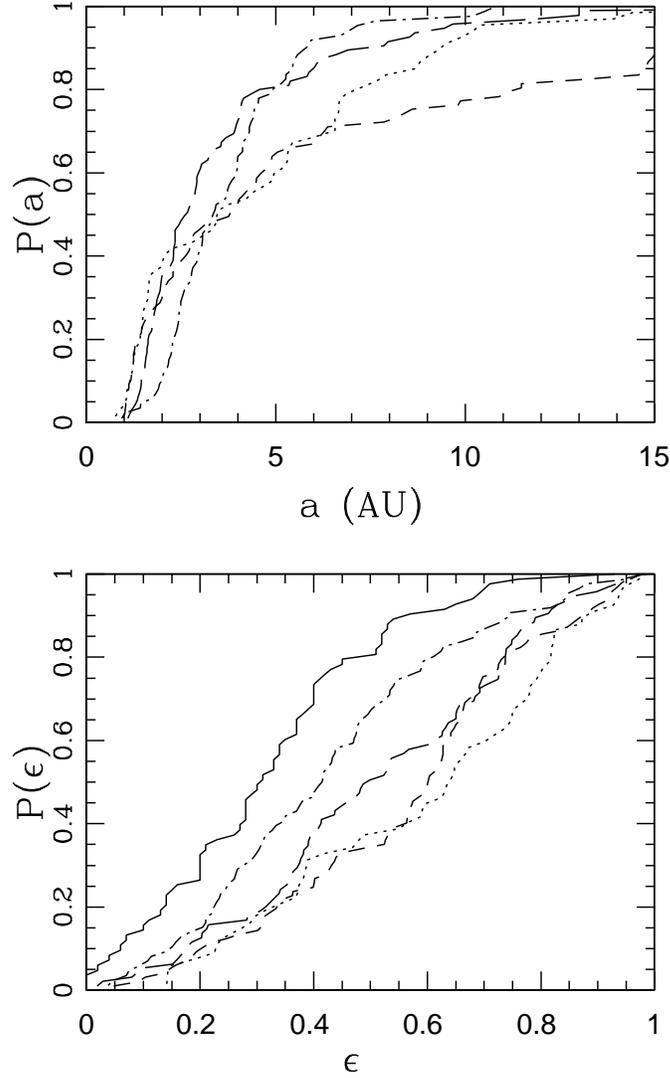}
\figcaption{Orbital properties of the innermost planet. The upper
panel shows the cumulative distribution for the semi-major axis of the
innermost surviving planet. The dotted curve represents planets with
$m_P$ = 1 $m_J$; the dashed curve represents planets with $m_P$ = 2
$m_J$; the long-dashed curve represents planets with randomly chosen
masses in the range $0 \le m_P \le 4 m_J$; and dot-dashed curve
represents planets with masses randomly chosen in $\log m$ in the
range $-1 \le \log_{10} [m_P/m_J] \le 1$. The lower panel shows the
corresponding distributions for the eccentricity of the innermost
planet. Also shown is the cumulative disitrubition of eccentricity for
the observed extra-solar planets (the subset of the sample with 
$a \le 0.1$ AU). }
\end{figure}

\newpage 
\begin{figure}
\figurenum{8}
\epsscale{1.0}
\plotone{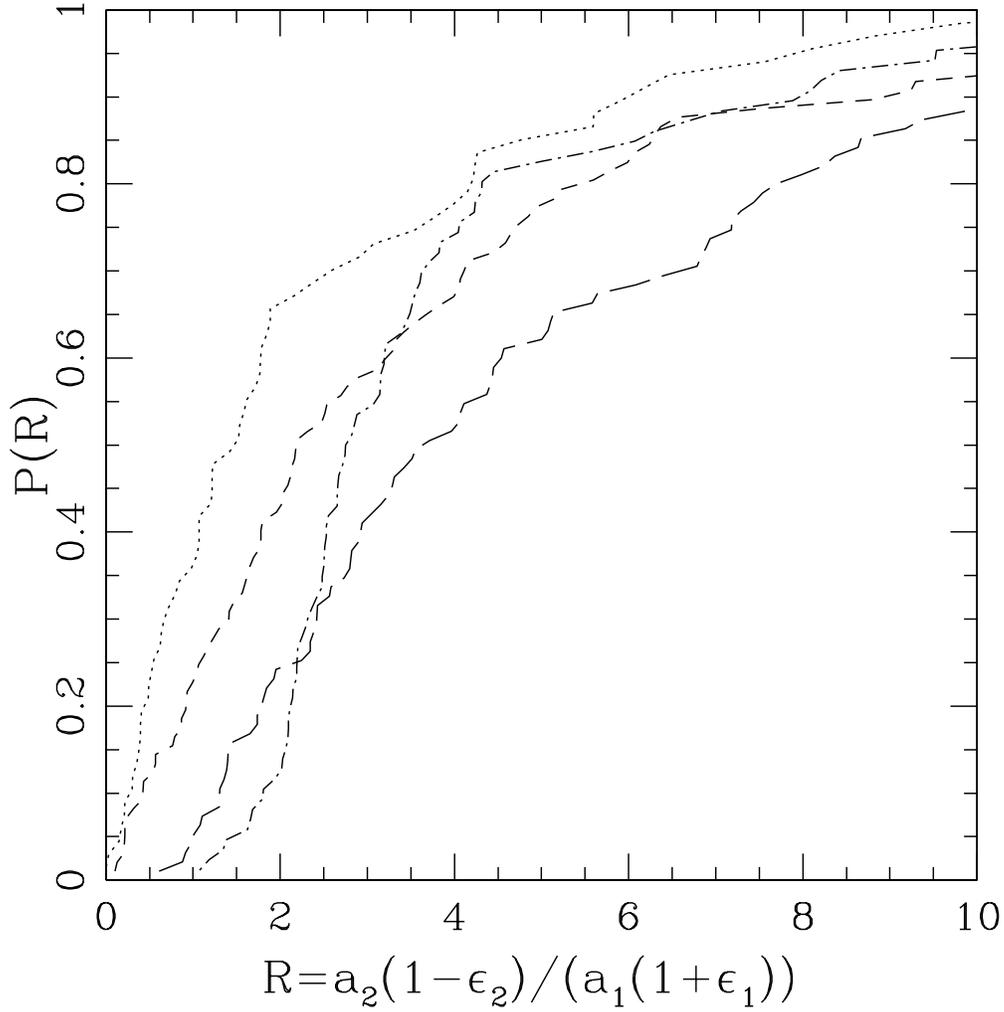}
\figcaption{The cumulative distribution of for the ratio 
$R$ = $a_2 (1 - \epsilon_2)/(a_1 (1 +  \epsilon_1))$, i.e., 
the ratio of periastron of the second surviving planet to the 
apastron of the innermost surviving planet. The dotted curve
represents planets with $m_P$ = 1 $m_J$; the dashed curve represents
planets with $m_P$ = 2 $m_J$; the long-dashed curve represents planets
with randomly chosen masses in the range $0 \le m_P \le 4 m_J$; and
dot-dashed curve represents planets with masses randomly chosen in
$\log m$ in the range $-1 \le \log_{10} [m_P/m_J] \le 1$. }
\end{figure}

\newpage 
\begin{figure}
\figurenum{9}
\epsscale{1.0}
\plotone{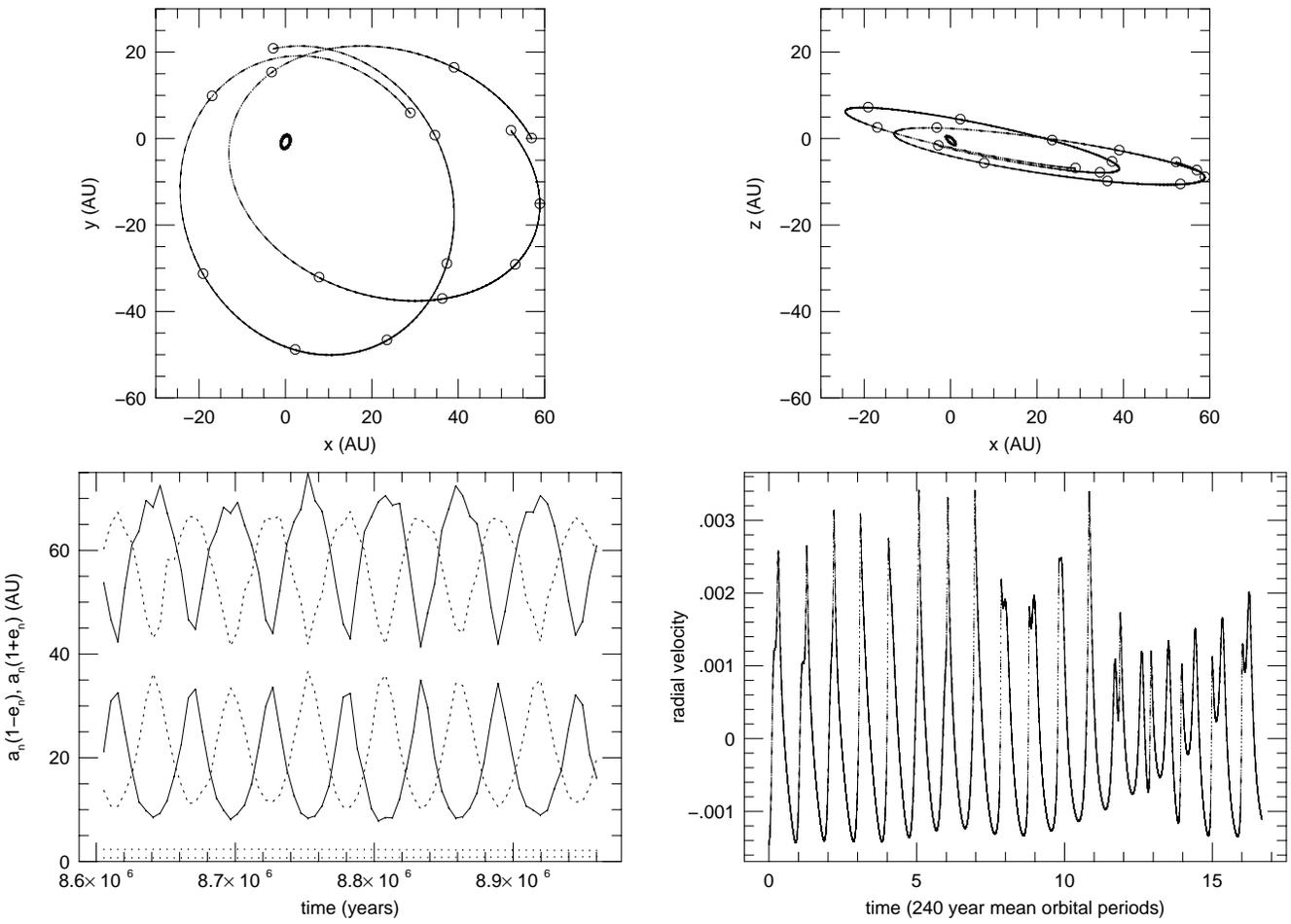}
\figcaption{The eccentric 1:1 resonance. The scattering experiments
often leave planets in resonant configurations, e.g., the 1:1
resonance depicted here. The lower left panel shows the history of the
system during the final million years of the simulation by plotting
$a(1-\epsilon)$ and $a(1+\epsilon)$ for each of surviving planets.
Sample trajectories of the planets are plotted in the upper two panels
(the eight fiducial points on each orbit show the planet locations
separated by 240/7=34.3 years). A sample radial velocity curve of the
star -- over 15 orbital periods -- is shown at the lower right. }
\end{figure}

\newpage 
\begin{figure}
\figurenum{10}
\epsscale{1.0}
\plotone{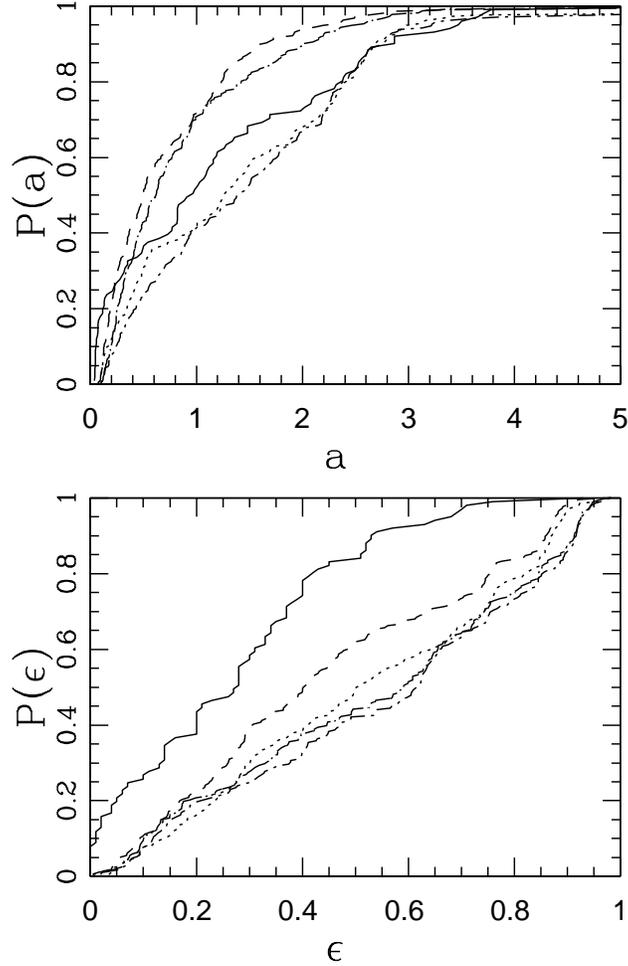}
\figcaption{Distributions of orbital properties for the surviving
planets in solar systems starting with two planets and a circumstellar
disk that exerts tidal torques on the outer planet.  The upper panel
shows the cumulative distributions for the semi-major axis of the
surviving planets.  The lower panel shows the corresponding cumulative
distributions for the eccentricity of the surviving planets. The solid 
curve depicts the distributions of the observed extrasolar planets. The 
distributions for the $\tau_{\rm ed}$ = 1 Myr simulations are shown as 
the long-dashed curves and the dotted curves (no evolution after the
first planet is lost).  The distributions for the $\tau_{\rm ed}$ = 3
Myr simulations are shown as the dot-dashed curves and the dashed
curves (no evolution after the first planet is lost). }

\end{figure} 

\newpage 
\begin{figure}
\figurenum{11}
\epsscale{1.0}
\plotone{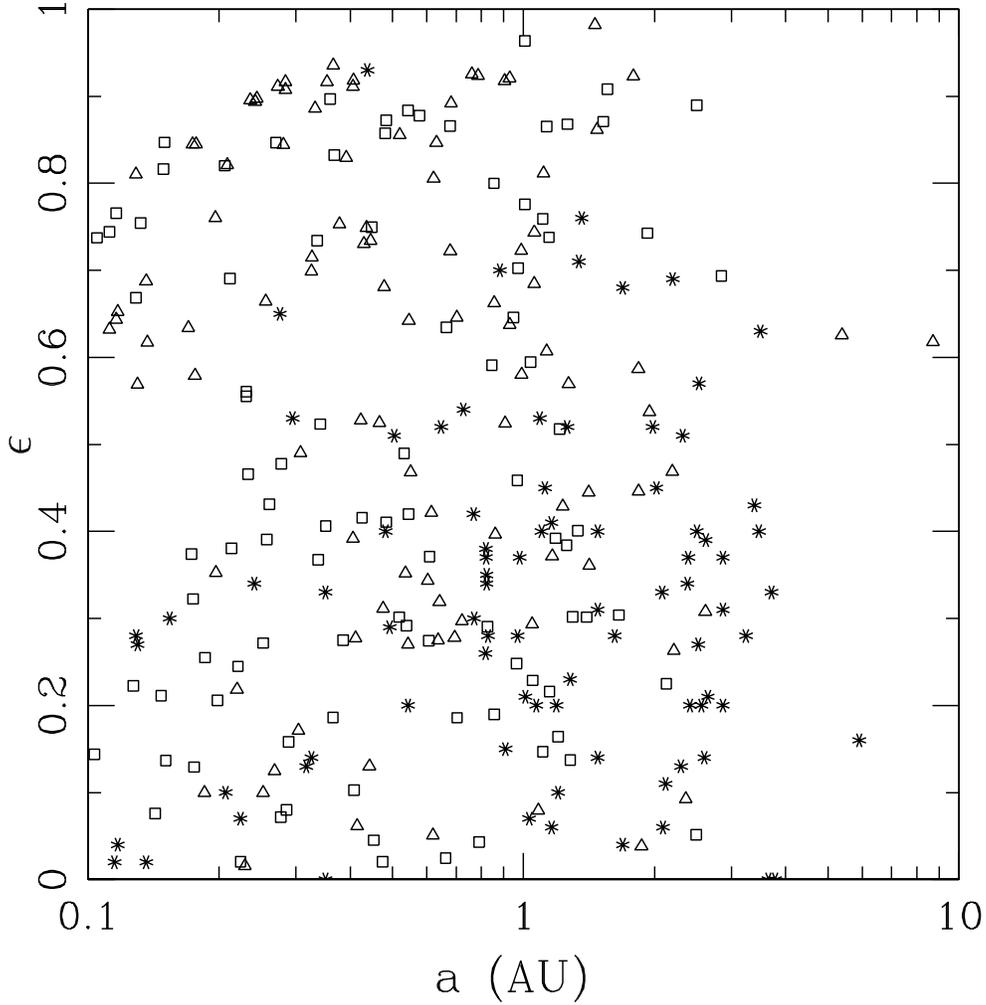}
\figcaption{The $a-\epsilon$ plane for the observed population of
extra-solar planets and the surviving planets in solar systems starting
with two planets surrounded by a circumstellar disk. The star symbols
represent the observed planetary orbits. The open triangles show the 
surviving planets for theoretical simulations using an eccentricity 
damping time scale $\tau_{\rm ed}$ = 3 Myr. The open squares show the 
surviving planets for simulations using $\tau_{\rm ed}$ = 1 Myr. } 
\end{figure} 

\end{document}